\documentclass[twocolumn]{revtex4}
\usepackage{color}
\def\gr{$\gamma$-ray}
\usepackage{aas_macros}
\usepackage{graphicx}

\begin{document}

\title{Neutrino signal from {a population of} Seyfert galaxies }
\author{A.~Neronov$^{1,2}$}
\author{D.~Savchenko$^{1,3,4}$}
\author{D.~V.~Semikoz$^{1}$}

\affiliation{$^{1}$Universit\'e Paris Cit\'e, CNRS, Astroparticule et Cosmologie, 
F-75013 Paris, France}

\affiliation{$^{2}$Laboratory of Astrophysics, \'Ecole Polytechnique F\'ed\'erale de Lausanne, CH-1015 Lausanne, Switzerland}

\affiliation{$^{3}$Bogolyubov Institute for Theoretical Physics of the NAS of Ukraine, 03143 Kyiv, Ukraine}

\affiliation{$^{4}$Kyiv Academic University, 03142 Kyiv, Ukraine}

\begin{abstract}
    IceCube collaboration has previously reported an evidence for neutrino signal from a Seyfert galaxy NGC 1068. This may suggest that all Seyfert galaxies emit neutrinos. To test this hypothesis, we identify the best candidate neutrino sources among nearby Seyfert galaxies, based on their hard X-ray properties. Only two other sources, NGC 4151 and NGC 3079 are expected to be detectable in 10 years of IceCube data. We  find an evidence {($\sim 3\sigma$)} for neutrino signal from both sources in publicly available ten-year IceCube dataset. {Though neither source alone is above the threshold for discovery,} the chance coincidence probability to find the observed neutrino count excesses in the directions of the two out of two expected sources, in addition to the previously reported brightest source,  is {$p<2.6\times 10^{-7}$}. This corresponds to a correlation between Seyfert galaxies and neutrino emission. 
\end{abstract}
\maketitle

\textbf{Introduction.}
Seyfert galaxies form the most abundant type of Active Galactic Nuclei (AGN) in the local Universe \cite{1943ApJ....97...28S,2022A&A...659A..32P}. Their emission spectra are dominated by the infrared-visible continuum originating from black hole accretion disk emission scattered by a dusty torus and by hard X-ray band emission from hot corona near the black hole. Most Seyfert galaxies are radio quiet sources, showing no or weak particle acceleration activity.  Recent high-resolution imaging of nearby Seyfert galaxies by Very Large Baseline Array (VLBA) \cite{2021ApJ...906...88F} shows that in most of the sources, weak radio emission originates from extended structures in the host galaxy, rather from the AGN itself. Only several sources show signatures of non-thermal synchrotron emission from the nucleus that may be related to a weak jet. Seyfert type AGN are also not strong \gr\ sources. Only two Seyfert galaxies, NGC 1068 and NGC 4945, are detected by Fermi Large Area Telescope (LAT) in the energy range above 100~MeV   \cite{2011ApJ...742...66T} and it is probable that the \gr\ emission is not related to the activity of the AGN, but may rather originate from starburst activity also found in these sources.  

In this respect, a recent report by IceCube collaboration presenting evidence for the neutrino signal from a Seyfert galaxy NGC 1068 \cite{2022Sci...378..538I} appears surprising and deserves larger scrutiny. Remarkably, the reported neutrino flux from the source in the TeV energy range is two orders of magnitude higher than the \gr\ flux at the same energy, derived from MAGIC observations of the source \cite{2019ApJ...883..135A}. This may possibly be explained by the compactness of the source. If the signal originates from the vicinity of the black hole, \gr s that are produced together with neutrinos in interactions of high-energy protons would not be able to leave the source because of the pair production on low-energy photons originating from the accretion disk and hot corona.  

It is not clear a-priori, if Seyfert galaxies are generically capable to accelerate protons to multi-TeV range and emit neutrinos or NGC 1068 is a peculiar source and neutrino emission from the source is not related to the Seyfert activity powered by the accretion on the supermassive black hole. It is possible that protons are accelerated in the accretion flows of all Seyfert galaxies \cite{Murase:2019vdl,Kheirandish:2021wkm} or in the black hole magnetosphere \cite{Neronov:2009zz,Aleksic:2014xsg} and is possibly related to the generation of jet, so that only jet-emitting Seyfert galaxies are neutrino sources. Finally, it may be that a specific event in NGC 1068 that is not typical to the entire Seyfert population is responsible for the neutrino emission. 

In what follows, we explore the possibility that Seyfert galaxies are generically sources of high-energy neutrinos. To test this hypothesis, we identify the best neutrino source candidates among nearby Seyfert galaxies and search for the neutrino signal from these best candidates using publicly available 10-year point source analysis dataset of IceCube \cite{icecube_10yr_paper}. 

\textbf{Neutrino -- hard X-ray flux scaling.}
We assume that the neutrino signal from NGC 1068 is a template for a typical Seyfert galaxy, and we want to find which other Seyfert galaxies should be detectable under this hypothesis. Neutrinos from high-energy proton interactions are produced together with \gr s, electrons and positrons.  Total power injected into the electromagnetic channel (electrons, positrons, \gr s) is comparable to the power injected into neutrinos. 

The electromagnetic power can be only dissipated radiatively. This means that the electromagnetic luminosity generated by high-energy proton interactions has to be of the same order as the neutrino luminosity of the source. The energies of \gr s, electrons and positrons produced in $pp$ and $p\gamma$ interactions are comparable to the energies of neutrinos.  The absence of \gr\ emission with the flux comparable to the neutrino flux from NGC 1068 at the energy $E_\gamma\sim 1$~TeV  indicates that the source is not transparent to \gr s \cite{Murase:2022dog}. \gr s  produce pairs in interactions with photons of energy $\epsilon_{ph}\simeq 1\left[E_\gamma/1\mbox{ TeV}\right]^{-1}\mbox{ eV}$. Such photons are emitted by the accretion flow that in NGC 1068 has the luminosity $L_{a}\sim 10^{45}$~erg/s \cite{2008A&A...485...33H}. The conventional accretion disk spectrum is $L_{a}(E)\sim EdL_{acc}/dE\propto E^{4/3}\exp(-E/E_{cut})$ that peaks at $E_{cut}\sim 100$~eV for the disk around a black hole of the mass $M\sim 10^7M_\odot$  \cite{Abramowicz:2011xu}. The strongest opacity for the $\gamma\gamma$ pair production is thus expected for the \gr s with energies $E_\gamma\sim 10$~GeV, with the optical depth
$\tau_{a}=\sigma_{\gamma\gamma}n_{ph}R\simeq 
10^6 \left[L_{a}/10^{45}\mbox{ erg/s}\right]\left[R/3\times 10^{12}\mbox{ cm}\right]^{-1}\left[E_\gamma/10\mbox{ GeV}\right]$
where $\sigma_{\gamma\gamma}\simeq 10^{-25}$~cm$^2$ is the pair production cross-section,
$
n_{ph}=L_{a}/(4\pi R^2 \epsilon_{ph}c)
$
is the density of soft photons and $R$ is the source size, comparable to the Schwarzschild radius $R\sim R_{schw}\simeq 3\times 10^{12}[M/10^7M_\odot]$~cm (for emission at $\epsilon_{ph}\sim 100$~eV). Assuming the radial accretion flow  temperature and luminosity dependence $T\propto R^{-4/3}$, $L_{a}\propto R^{-1}$, one can estimate the energy dependence of the optical depth
$\tau_{a}\sim 10^6\left(E/10\mbox{ GeV}\right)^{-3/2}$, so that the source may be opaque to \gr s with energies up to approximately 100~TeV.

Gamma-rays with energies below 10~GeV are absorbed in interactions with X-ray photons from hot corona with temperature reaching $T_c\sim 100$~keV.  Its optical depth for this process is  
$\tau_{c}\simeq 
 7 \left[L_{c}/10^{43}\mbox{ erg/s}\right]\left[R/3\times 10^{12}\mbox{ cm}\right]^{-1}\left[E_\gamma/10\mbox{ MeV}\right] $
where $L_c$ is the luminosity of the corona. Thus, most of the electromagnetic power from $pp$ and $p\gamma$ interactions, comparable to the neutrino luminosity of the source, has to be released in the energy range $E\lesssim 1$~MeV \cite{2022ApJ...939...43E}. A linear scaling between the neutrino and secondary hard X-ray / soft \gr\ flux from high-energy proton interactions is expected in this case. 

The TeV band muon neutrino luminosity of NGC 1068 is estimated as \cite{2022Sci...378..538I}
$L_{\nu_\mu}\sim 4\pi D^2F_{\nu_\mu}\sim 2\times 10^{42}\left[F_{\nu_\mu}/5\times 10^{-11}\mbox{ TeV/cm}^{2}\mbox{s}\right]\mbox{erg}/\mbox{s}$ where $D\simeq 16.3$~Mpc is the distance to the source. This is approximately $L_{\nu_\mu,\ TeV}\sim 0.02 L_{hX0}$ of the intrinsic hard X-ray band source luminosity  $L_{hX0}$ in the hard X-ray band. NGC 1068  is a Compton-thick AGN, with X-ray flux attenuated by the Compton scattering through a medium with the column density \cite{2000MNRAS.318..173M}  $N_{H} \gtrsim 10^{25}\mbox{ cm}^2$.
The hard X-ray flux arriving at Earth is  $F_{hX}\sim L_{hX0}/(4\pi D^2)\exp(-\tau_C)$ where the optical depth for the Compton scattering is  $\tau_C=\sigma_T N_H\simeq 7\left[N_H/10^{25}\mbox{ cm}^{-2}\right]$.
Apart from the attenuated flux from the corona, the hard X-ray flux has a contribution from Compton reflection that may even dominate the observed flux for heavily obscured sources, like NGC 1068. This introduces large uncertainty in the  estimates of the intrinsic luminosity of the corona for such sources \cite{2016MNRAS.456L..94M}. The power released by high-energy proton interactions contributes to the intrinsic hard X-ray luminosity of the source and hence the neutrino luminosity is expected to scale with the intrinsic, rather than observed, hard X-ray luminosity in Compton-thick sources of Seyfert 2 type. 

\textbf{Source selection.}
The linear scaling of neutrino and  the secondary sub-MeV electromagnetic luminosity from the power released in interactions of high-energy protons suggests that Seyfert galaxies with the highest unabsorbed sub-MeV flux should be the brightest neutrino sources. 

To define {a pre-determined (a-priori)} neutrino source candidate catalog,  we follow the approach of  Ref. \cite{2021ApJ...906...88F} and start from a volume-complete sample of nearby Seyfert galaxies above the luminosity threshold $L_{min}=10^{42}$~erg/s from the Swift-BAT 105 months survey \cite{2018ApJS..235....4O}. We consider sources in the declination range $-5^\circ<\delta<60^\circ$ in which IceCube can observe in the muon neutrino channel at moderate atmospheric background levels and without strong absorption by the Earth.  We include in our candidate  list sources that are confirmed Seyfert galaxies, based on the Turin Seyfert galaxy catalog \cite{2022A&A...659A..32P}. 

This pre-selects 13 sources in the sky region of interest, listed in Table I of the {Supplementary Material  that includes information on the sources, further description of the selection criteria and References \cite{IceCube:2019cia,IceCube:2018cha,IceCube:2022der,2019...621A..28G,2014ApJ...794..111B,2020...640A..31P,2018ApJ...854...49M,Bauer:2014rla,2011PhRvD..83a2001A,1997ASPC..121..101M,2010MNRAS.402..724P}.}  Apart from NGC 1068, three other sources, NGC 1320, NGC 3079 and NGC 7479 are Compton-thick and two other, NGC 4388 and NGC 5899, have $N_H$ in excess of $10^{23}$~cm$^{2}$.  For these sources, we find the estimates of the intrinsic hard X-ray luminosity based on the detailed modelling of the  spectra measured by  NUSTAR telescope (with higher signal-to-noise compared to Swift-BAT) reported in the literature \cite{2016MNRAS.456L..94M,2017PhDT........40B,2022ApJ...936..149P}.

\begin{figure}
    \includegraphics[width=\linewidth]{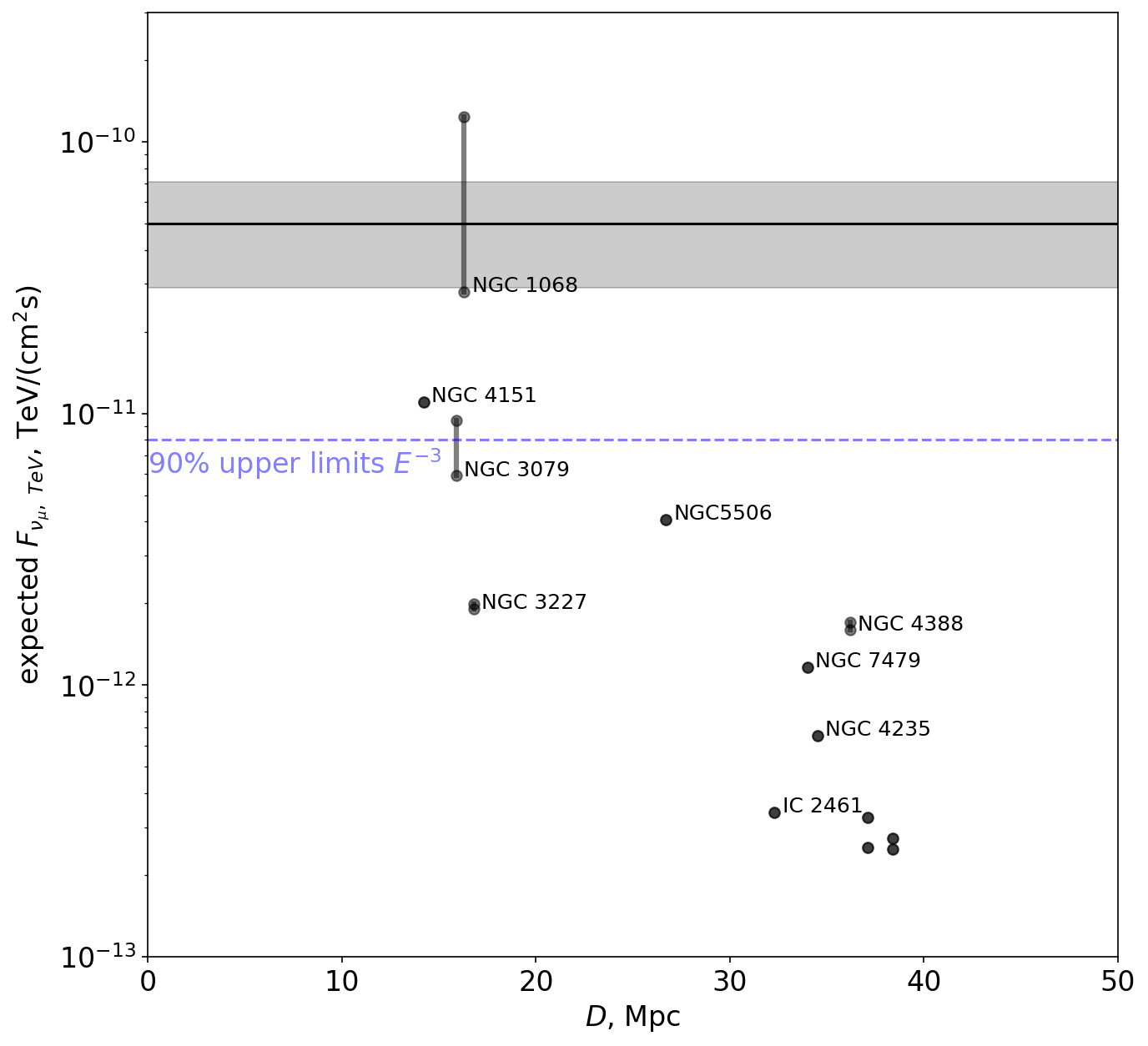}
    \caption{Expected neutrino fluxes of Seyfert galaxies derived from the hard X-ray data. Vertical lines correspond to the uncertainty of the  intrinsic hard X-ray flux estimates for Compton-thick sources. Horizontal black line and grey band show the measured neutrino flux of NGC 1068 \cite{2022Sci...378..538I}. Horizontal dashed  line shows the expected level of 90\% upper limits on neutrino flux for sources with $E^{-3}$ powerlaw spectra in the declination range $-5^\circ<$ to $<60^\circ$, attainable with 10-year IceCube exposure \cite{icecube_10yr_paper}. }
    \label{fig:z_FhX}
\end{figure}

Fig. \ref{fig:z_FhX} shows the estimates of the muon neutrino fluxes  from the pre-selected sources. Horizontal dashed line shows the sensitivity limit of the 10-year IceCube exposure for the $E^{-3}$ powerlaw neutrino spectrum (with the slope close to the measured slope of the NGC 1068 spectrum \cite{2022Sci...378..538I}). Only two additional sources may be detectable individually in the 10-year IceCube exposure: NGC 4151, a Seyfert 1 galaxy, and a Seyfert 2 galaxy, NGC 3079. NGC 3079  is  similar to NGC 1068 in the sense that it is also Compton-thick with $N_H>10^{24}$~cm$^{-2}$. For NGC 4151 the column density is much lower. 

The requirement that the source should be above the sensitivity limit of IceCube reduces the number of sources from 13 to 3. We exclude NGC 1068 from our final source catalog, because this source has been used to formulate the hypothesis on the scaling of the neutrino flux with the hard X-ray intrinsic source luminosity. Thus, the final "neutrino candidate" source catalog consists of only two sources.

\textbf{IceCube data analysis}.
We search for the neutrino signal from these two potentially detectable sources using publicly available 10-year dataset of IceCube \cite{2021arXiv210109836I}. Similar to  \cite{2022Sci...378..538I}, we consider only the data of the fully assembled 86 string detector that has homogeneous event selection and stable instrument response functions.  We perform the unbinned likelihood analysis \cite{mattox96}, see Supplementary Material for details.

\begin{figure}
    \includegraphics[width=\linewidth]{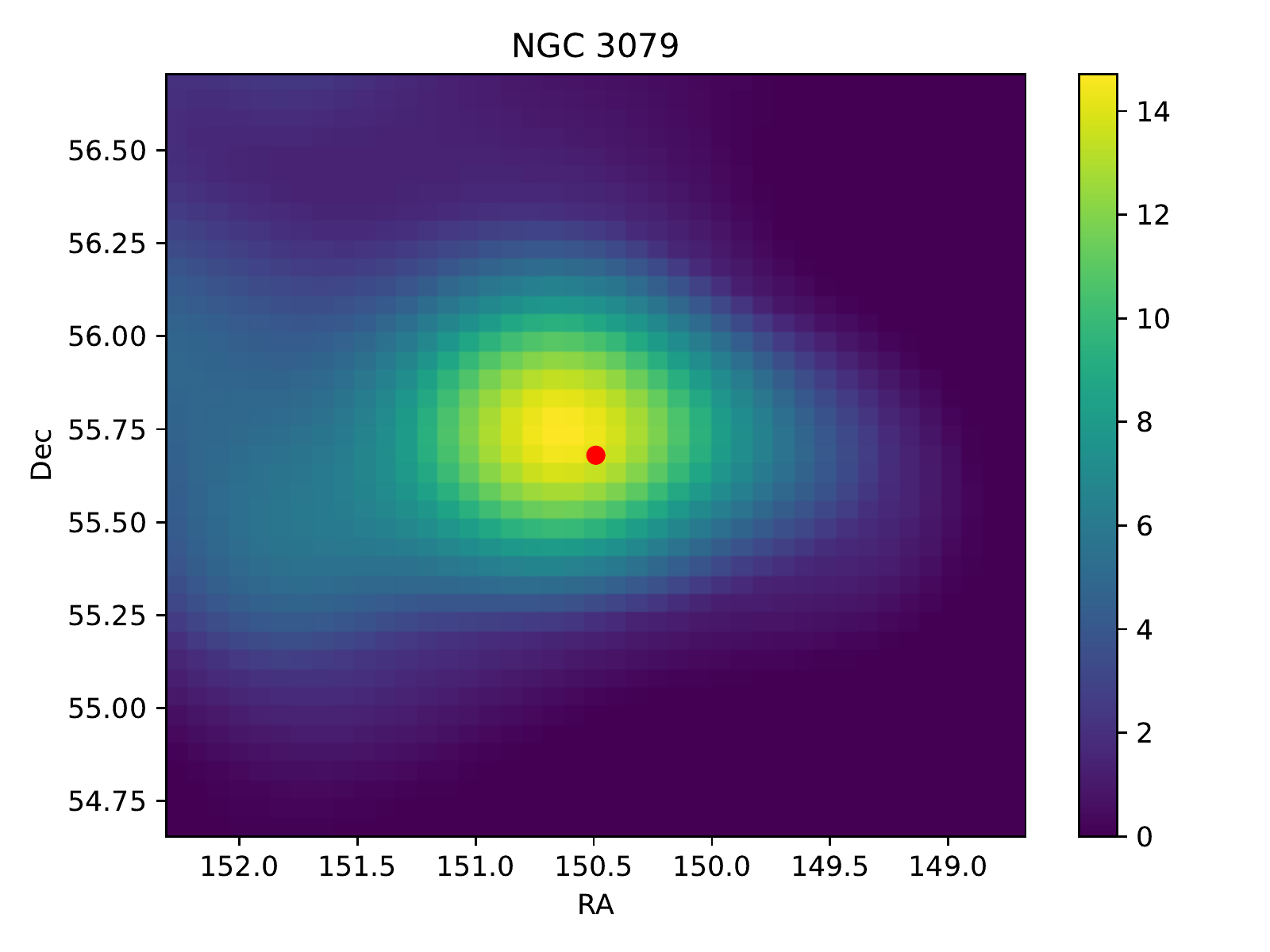}
    \includegraphics[width=\linewidth]{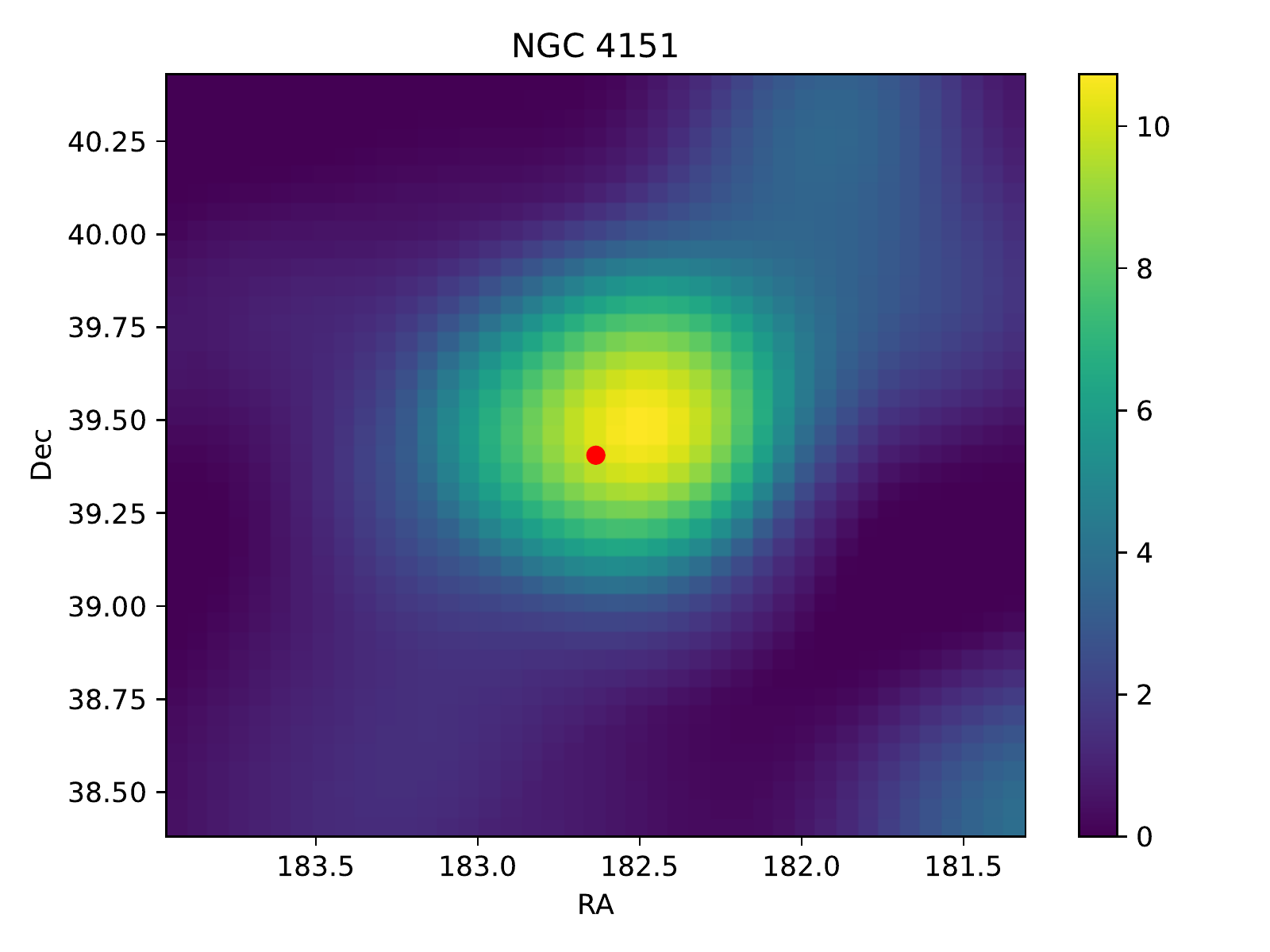}
    \caption{Maps of Test Statistic values around the positions of NGC 3079 and NGC 4151. Red dots mark the catalog source positions.}
    \label{fig:TS_map_3079}
\end{figure}

\begin{figure}
    \includegraphics[width=\linewidth]{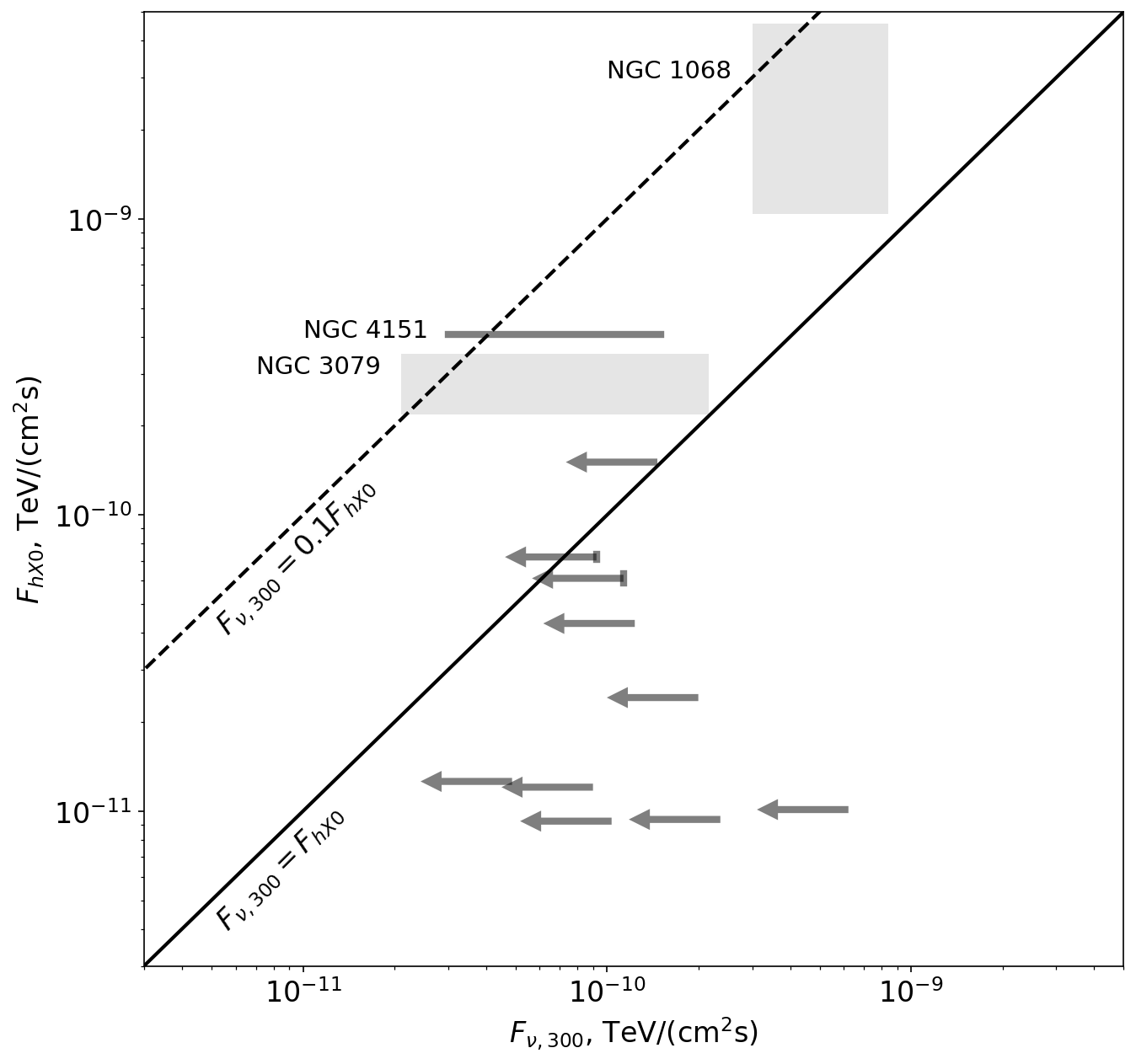}
    \caption{Comparison of the intrinsic hard X-ray fluxes (15-195 keV range) with the all-flavour neutrino flux measurements or upper limits in the energy range above 300~GeV (see Supplementary Material) . Grey boxes correspond to the uncertainties of the flux for detected sources.  Black solid line shows the $F_{\nu,300}=F_{hX0}$  scaling, dashed line is for the neutrino flux ten times smaller than the hard X-ray flux. }
    \label{fig:X-ray_nu}
\end{figure}

Fig. \ref{fig:TS_map_3079} shows the map of Test Statistic  values around the positions of NGC 3079 and NGC 4151. For each source, an evidence for the signal is found in the data. In the case of NGC 3079, the maximal Test Statistic value is found at the Right Ascension  $150.7^\circ$, Declination $55.7^\circ$. The Test Statistic value at the catalog source position is $14.1$. The probability that this or higher Test Statistic value is found in a background fluctuation is $p_{3079}=9.3\times 10^{-5}$. The $0.3-100$~TeV flux is $F_{\nu_\mu,100}=3.2+4.0-2.5\times10^{-11}\mathrm{\frac{TeV}{cm^2s}}$. For NGC 4151,  the excess  at the source position has the Test Statistic value $10.0$. Such an  excess can be found in background fluctuations with probability $p_{4151}=2.7\times 10^{-3}$.  The highest Test Statistic is found at the position Right Ascension $182.5^\circ$ and Declination $39.5^\circ$, just $0.1^\circ$ from the catalog source position. The $0.3-100$~TeV flux is estimated to be $F_{\nu_\mu,100}=2.8+2.2-2.0\times10^{-11}\mathrm{\frac{TeV}{cm^2s}}$. Overall, two out of two additional sources show an evidence for the signal in the IceCube data.  The probability to find  random background count fluctuations at the two positions is  $p=p_{3079}p_{4151}\simeq 2.6\times 10^{-7}$.  No other Seyfert galaxy from our source sample shows an excess in our analysis. Fig. \ref{fig:X-ray_nu} shows a comparison between neutrino flux estimates and upper limits and  hard X-ray corona fluxes of the selected sources.

\textbf{Discussion.}
Analysis of the IceCube data presented above {reveals a correlation between Seyfert galaxies and neutrino emission.} Estimates of the neutrino flux based on the hard X-ray luminosity of the central engines of Seyfert type AGN has suggested that only two additional sources, besides NGC 1068, should have been detected in the 10-year IceCube data sample. We have found excess neutrino counts at the positions of both additional sources, NGC 3079 and NGC 4151. The chance coincidence probability to find the observed excess in both sources is $p\simeq 2.6\times 10^{-7}$, which corresponds to the $5\sigma$ confidence level detection of {the correlation}. 

The spectra of neutrino emission from the three sources  are softer than $E^{-2}$, (See Supplementary Material). This means that most of the neutrino power is emitted in the energy range close to the energy threshold of IceCube (at several hundred GeV). Fig. \ref{fig:X-ray_nu} shows a comparison of the integral all-flavour neutrino flux in the IceCube energy band (we assume equal flux per neutrino flavour) with the hot corona flux in the Swift/BAT energy range that would be observed if the sources would be Compton-thin.  One can see that the neutrino source power is comparable to the power emitted by the  hot corona. This suggests that interactions of high-energy protons may be an important element of the energy balance of the hot corona around the black hole accretion disk. This fact may call for a revision of the models of the hot coronae \cite{galaxies6020044}. 

Seyfert galaxies provide a major contribution to the X-ray background that peaks at the energy $\sim 30$~keV at the flux level $F_{XRB}\sim 10^{-4.5}$~GeV/(cm$^2$s sr) \cite{Churazov:2006bk}. If all Seyfert galaxies {would} emit neutrinos with the power at the level of 0.1-1 of their hard X-ray power, and with soft neutrino spectrum with slope $\Gamma>2$ extending into (or below) the 100~GeV range, the cumulative neutrino flux from the Seyfert galaxy population {would}  be at the level of 0.1-1 of hard X-ray background level.  In this case,  Seyfert galaxies may also provide a major contribution to the observed astrophysical neutrino flux \cite{IceCube:2020acn}.  

The origin of the high-energy protons whose interactions lead to neutrino emission from the cores of Seyfert galaxies is not clear. {Difference} between the neutrino and \gr\ luminosity in the TeV energy range suggests compact size of the neutrino source and locates proton acceleration site close to the black hole. NGC 1068, NGC 3079 and NGC 4151 all have detectable radio flux from the central parsec around the black hole, revealed by the VLBA detections \cite{2021ApJ...906...88F}. This radio emission may be associated to particle acceleration activity close to black hole. However, the angular resolution of radio observations is not sufficient for localising the acceleration site. It may be that the acceleration happens at the base of a weak jet ejected by the black hole. Alternatively, reconnection in the accretion disk or a vacuum gap in the black hole magnetosphere may be considered as candidate proton acceleration sites \cite{Neronov:2009zz,Aleksic:2014xsg}.

\bibliography{references}

\begin{thebibliography}{36}
\expandafter\ifx\csname natexlab\endcsname\relax\def\natexlab#1{#1}\fi
\expandafter\ifx\csname bibnamefont\endcsname\relax
  \def\bibnamefont#1{#1}\fi
\expandafter\ifx\csname bibfnamefont\endcsname\relax
  \def\bibfnamefont#1{#1}\fi
\expandafter\ifx\csname citenamefont\endcsname\relax
  \def\citenamefont#1{#1}\fi
\expandafter\ifx\csname url\endcsname\relax
  \def\url#1{\texttt{#1}}\fi
\expandafter\ifx\csname urlprefix\endcsname\relax\def\urlprefix{URL }\fi
\providecommand{\bibinfo}[2]{#2}
\providecommand{\eprint}[2][]{\url{#2}}

\bibitem[{\citenamefont{{Seyfert}}(1943)}]{1943ApJ....97...28S}
\bibinfo{author}{\bibfnamefont{C.~K.} \bibnamefont{{Seyfert}}},
  \bibinfo{journal}{\apj} \textbf{\bibinfo{volume}{97}}, \bibinfo{pages}{28}
  (\bibinfo{year}{1943}).

\bibitem[{\citenamefont{{Pe{\~n}a-Herazo}
  et~al.}(2022)\citenamefont{{Pe{\~n}a-Herazo}, {Massaro}, {Chavushyan},
  {Masetti}, {Paggi}, and {Capetti}}}]{2022A&A...659A..32P}
\bibinfo{author}{\bibfnamefont{H.~A.} \bibnamefont{{Pe{\~n}a-Herazo}}},
  \bibinfo{author}{\bibfnamefont{F.}~\bibnamefont{{Massaro}}},
  \bibinfo{author}{\bibfnamefont{V.}~\bibnamefont{{Chavushyan}}},
  \bibinfo{author}{\bibfnamefont{N.}~\bibnamefont{{Masetti}}},
  \bibinfo{author}{\bibfnamefont{A.}~\bibnamefont{{Paggi}}}, \bibnamefont{and}
  \bibinfo{author}{\bibfnamefont{A.}~\bibnamefont{{Capetti}}},
  \bibinfo{journal}{\aap} \textbf{\bibinfo{volume}{659}}, \bibinfo{eid}{A32}
  (\bibinfo{year}{2022}).

\bibitem[{\citenamefont{{Fischer} et~al.}(2021)\citenamefont{{Fischer},
  {Secrest}, {Johnson}, {Dorland}, {Cigan}, {Fernandez}, {Hunt}, {Koss},
  {Schmitt}, and {Zacharias}}}]{2021ApJ...906...88F}
\bibinfo{author}{\bibfnamefont{T.~C.} \bibnamefont{{Fischer}}},
  \bibinfo{author}{\bibfnamefont{N.~J.} \bibnamefont{{Secrest}}},
  \bibinfo{author}{\bibfnamefont{M.~C.} \bibnamefont{{Johnson}}},
  \bibinfo{author}{\bibfnamefont{B.~N.} \bibnamefont{{Dorland}}},
  \bibinfo{author}{\bibfnamefont{P.~J.} \bibnamefont{{Cigan}}},
  \bibinfo{author}{\bibfnamefont{L.~C.} \bibnamefont{{Fernandez}}},
  \bibinfo{author}{\bibfnamefont{L.~R.} \bibnamefont{{Hunt}}},
  \bibinfo{author}{\bibfnamefont{M.}~\bibnamefont{{Koss}}},
  \bibinfo{author}{\bibfnamefont{H.~R.} \bibnamefont{{Schmitt}}},
  \bibnamefont{and}
  \bibinfo{author}{\bibfnamefont{N.}~\bibnamefont{{Zacharias}}},
  \bibinfo{journal}{\apj} \textbf{\bibinfo{volume}{906}}, \bibinfo{eid}{88}
  (\bibinfo{year}{2021}), \eprint{2011.06570}.

\bibitem[{\citenamefont{{Teng} et~al.}(2011)\citenamefont{{Teng}, {Mushotzky},
  {Sambruna}, {Davis}, and {Reynolds}}}]{2011ApJ...742...66T}
\bibinfo{author}{\bibfnamefont{S.~H.} \bibnamefont{{Teng}}},
  \bibinfo{author}{\bibfnamefont{R.~F.} \bibnamefont{{Mushotzky}}},
  \bibinfo{author}{\bibfnamefont{R.~M.} \bibnamefont{{Sambruna}}},
  \bibinfo{author}{\bibfnamefont{D.~S.} \bibnamefont{{Davis}}},
  \bibnamefont{and} \bibinfo{author}{\bibfnamefont{C.~S.}
  \bibnamefont{{Reynolds}}}, \bibinfo{journal}{\apj}
  \textbf{\bibinfo{volume}{742}}, \bibinfo{eid}{66} (\bibinfo{year}{2011}),
  \eprint{1109.2734}.

\bibitem[{\citenamefont{{IceCube Collaboration}
  et~al.}(2022)\citenamefont{{IceCube Collaboration}, {Abbasi}, {Ackermann},
  {Adams}, {Aguilar}, {Ahlers}, {Ahrens}, {Alameddine}, {Alispach}, {Alves}
  et~al.}}]{2022Sci...378..538I}
\bibinfo{author}{\bibnamefont{{IceCube Collaboration}}},
  \bibinfo{author}{\bibfnamefont{R.}~\bibnamefont{{Abbasi}}},
  \bibinfo{author}{\bibfnamefont{M.}~\bibnamefont{{Ackermann}}},
  \bibinfo{author}{\bibfnamefont{J.}~\bibnamefont{{Adams}}},
  \bibinfo{author}{\bibfnamefont{J.~A.} \bibnamefont{{Aguilar}}},
  \bibinfo{author}{\bibfnamefont{M.}~\bibnamefont{{Ahlers}}},
  \bibinfo{author}{\bibfnamefont{M.}~\bibnamefont{{Ahrens}}},
  \bibinfo{author}{\bibfnamefont{J.~M.} \bibnamefont{{Alameddine}}},
  \bibinfo{author}{\bibfnamefont{C.}~\bibnamefont{{Alispach}}},
  \bibinfo{author}{\bibfnamefont{J.}~\bibnamefont{{Alves}},
  \bibfnamefont{A.~A.}}, \bibnamefont{et~al.}, \bibinfo{journal}{Science}
  \textbf{\bibinfo{volume}{378}}, \bibinfo{pages}{538} (\bibinfo{year}{2022}),
  \eprint{2211.09972}.

\bibitem[{\citenamefont{{Acciari} et~al.}(2019)\citenamefont{{Acciari},
  {Ansoldi}, {Antonelli}, {Arbet Engels}, {Baack}, {Babi{\'c}}, {Banerjee},
  {Barres de Almeida}, {Barrio}, {Becerra Gonz{\'a}lez}
  et~al.}}]{2019ApJ...883..135A}
\bibinfo{author}{\bibfnamefont{V.~A.} \bibnamefont{{Acciari}}},
  \bibinfo{author}{\bibfnamefont{S.}~\bibnamefont{{Ansoldi}}},
  \bibinfo{author}{\bibfnamefont{L.~A.} \bibnamefont{{Antonelli}}},
  \bibinfo{author}{\bibfnamefont{A.}~\bibnamefont{{Arbet Engels}}},
  \bibinfo{author}{\bibfnamefont{D.}~\bibnamefont{{Baack}}},
  \bibinfo{author}{\bibfnamefont{A.}~\bibnamefont{{Babi{\'c}}}},
  \bibinfo{author}{\bibfnamefont{B.}~\bibnamefont{{Banerjee}}},
  \bibinfo{author}{\bibfnamefont{U.}~\bibnamefont{{Barres de Almeida}}},
  \bibinfo{author}{\bibfnamefont{J.~A.} \bibnamefont{{Barrio}}},
  \bibinfo{author}{\bibfnamefont{J.}~\bibnamefont{{Becerra Gonz{\'a}lez}}},
  \bibnamefont{et~al.}, \bibinfo{journal}{\apj} \textbf{\bibinfo{volume}{883}},
  \bibinfo{eid}{135} (\bibinfo{year}{2019}), \eprint{1906.10954}.

\bibitem[{\citenamefont{Murase et~al.}(2020)\citenamefont{Murase, Kimura, and
  Meszaros}}]{Murase:2019vdl}
\bibinfo{author}{\bibfnamefont{K.}~\bibnamefont{Murase}},
  \bibinfo{author}{\bibfnamefont{S.~S.} \bibnamefont{Kimura}},
  \bibnamefont{and} \bibinfo{author}{\bibfnamefont{P.}~\bibnamefont{Meszaros}},
  \bibinfo{journal}{Phys. Rev. Lett.} \textbf{\bibinfo{volume}{125}},
  \bibinfo{pages}{011101} (\bibinfo{year}{2020}), \eprint{1904.04226}.

\bibitem[{\citenamefont{Kheirandish et~al.}(2021)\citenamefont{Kheirandish,
  Murase, and Kimura}}]{Kheirandish:2021wkm}
\bibinfo{author}{\bibfnamefont{A.}~\bibnamefont{Kheirandish}},
  \bibinfo{author}{\bibfnamefont{K.}~\bibnamefont{Murase}}, \bibnamefont{and}
  \bibinfo{author}{\bibfnamefont{S.~S.} \bibnamefont{Kimura}},
  \bibinfo{journal}{Astrophys. J.} \textbf{\bibinfo{volume}{922}},
  \bibinfo{pages}{45} (\bibinfo{year}{2021}), \eprint{2102.04475}.

\bibitem[{\citenamefont{Neronov et~al.}(2009)\citenamefont{Neronov, Semikoz,
  and Tkachev}}]{Neronov:2009zz}
\bibinfo{author}{\bibfnamefont{A.~Y.} \bibnamefont{Neronov}},
  \bibinfo{author}{\bibfnamefont{D.~V.} \bibnamefont{Semikoz}},
  \bibnamefont{and} \bibinfo{author}{\bibfnamefont{I.~I.}
  \bibnamefont{Tkachev}}, \bibinfo{journal}{New J. Phys.}
  \textbf{\bibinfo{volume}{11}}, \bibinfo{pages}{065015}
  (\bibinfo{year}{2009}), \eprint{0712.1737}.

\bibitem[{\citenamefont{Aleksic et~al.}(2014)}]{Aleksic:2014xsg}
\bibinfo{author}{\bibfnamefont{J.}~\bibnamefont{Aleksic}} \bibnamefont{et~al.},
  \bibinfo{journal}{Science} \textbf{\bibinfo{volume}{346}},
  \bibinfo{pages}{1080} (\bibinfo{year}{2014}), \eprint{1412.4936}.

\bibitem[{\citenamefont{{Aartsen} et~al.}(2020)\citenamefont{{Aartsen},
  {Ackermann}, {Adams}, {Aguilar}, {Ahlers}, {Ahrens}, {Alispach}, {Andeen},
  {Anderson}, {Ansseau} et~al.}}]{icecube_10yr_paper}
\bibinfo{author}{\bibfnamefont{M.~G.} \bibnamefont{{Aartsen}}},
  \bibinfo{author}{\bibfnamefont{M.}~\bibnamefont{{Ackermann}}},
  \bibinfo{author}{\bibfnamefont{J.}~\bibnamefont{{Adams}}},
  \bibinfo{author}{\bibfnamefont{J.~A.} \bibnamefont{{Aguilar}}},
  \bibinfo{author}{\bibfnamefont{M.}~\bibnamefont{{Ahlers}}},
  \bibinfo{author}{\bibfnamefont{M.}~\bibnamefont{{Ahrens}}},
  \bibinfo{author}{\bibfnamefont{C.}~\bibnamefont{{Alispach}}},
  \bibinfo{author}{\bibfnamefont{K.}~\bibnamefont{{Andeen}}},
  \bibinfo{author}{\bibfnamefont{T.}~\bibnamefont{{Anderson}}},
  \bibinfo{author}{\bibfnamefont{I.}~\bibnamefont{{Ansseau}}},
  \bibnamefont{et~al.}, \bibinfo{journal}{\prl} \textbf{\bibinfo{volume}{124}},
  \bibinfo{eid}{051103} (\bibinfo{year}{2020}), \eprint{1910.08488}.

\bibitem[{\citenamefont{Murase}(2022)}]{Murase:2022dog}
\bibinfo{author}{\bibfnamefont{K.}~\bibnamefont{Murase}},
  \bibinfo{journal}{Astrophys. J. Lett.} \textbf{\bibinfo{volume}{941}},
  \bibinfo{pages}{L17} (\bibinfo{year}{2022}), \eprint{2211.04460}.

\bibitem[{\citenamefont{{H{\"o}nig} et~al.}(2008)\citenamefont{{H{\"o}nig},
  {Prieto}, and {Beckert}}}]{2008A&A...485...33H}
\bibinfo{author}{\bibfnamefont{S.~F.} \bibnamefont{{H{\"o}nig}}},
  \bibinfo{author}{\bibfnamefont{M.~A.} \bibnamefont{{Prieto}}},
  \bibnamefont{and}
  \bibinfo{author}{\bibfnamefont{T.}~\bibnamefont{{Beckert}}},
  \bibinfo{journal}{\aap} \textbf{\bibinfo{volume}{485}}, \bibinfo{pages}{33}
  (\bibinfo{year}{2008}), \eprint{0804.0236}.

\bibitem[{\citenamefont{Abramowicz and Fragile}(2013)}]{Abramowicz:2011xu}
\bibinfo{author}{\bibfnamefont{M.~A.} \bibnamefont{Abramowicz}}
  \bibnamefont{and} \bibinfo{author}{\bibfnamefont{P.~C.}
  \bibnamefont{Fragile}}, \bibinfo{journal}{Living Rev. Rel.}
  \textbf{\bibinfo{volume}{16}}, \bibinfo{pages}{1} (\bibinfo{year}{2013}),
  \eprint{1104.5499}.

\bibitem[{\citenamefont{{Eichmann} et~al.}(2022)\citenamefont{{Eichmann},
  {Oikonomou}, {Salvatore}, {Dettmar}, and {Tjus}}}]{2022ApJ...939...43E}
\bibinfo{author}{\bibfnamefont{B.}~\bibnamefont{{Eichmann}}},
  \bibinfo{author}{\bibfnamefont{F.}~\bibnamefont{{Oikonomou}}},
  \bibinfo{author}{\bibfnamefont{S.}~\bibnamefont{{Salvatore}}},
  \bibinfo{author}{\bibfnamefont{R.-J.} \bibnamefont{{Dettmar}}},
  \bibnamefont{and} \bibinfo{author}{\bibfnamefont{J.~B.}
  \bibnamefont{{Tjus}}}, \bibinfo{journal}{\apj}
  \textbf{\bibinfo{volume}{939}}, \bibinfo{eid}{43} (\bibinfo{year}{2022}),
  \eprint{2207.00102}.

\bibitem[{\citenamefont{{Matt} et~al.}(2000)\citenamefont{{Matt}, {Fabian},
  {Guainazzi}, {Iwasawa}, {Bassani}, and {Malaguti}}}]{2000MNRAS.318..173M}
\bibinfo{author}{\bibfnamefont{G.}~\bibnamefont{{Matt}}},
  \bibinfo{author}{\bibfnamefont{A.~C.} \bibnamefont{{Fabian}}},
  \bibinfo{author}{\bibfnamefont{M.}~\bibnamefont{{Guainazzi}}},
  \bibinfo{author}{\bibfnamefont{K.}~\bibnamefont{{Iwasawa}}},
  \bibinfo{author}{\bibfnamefont{L.}~\bibnamefont{{Bassani}}},
  \bibnamefont{and}
  \bibinfo{author}{\bibfnamefont{G.}~\bibnamefont{{Malaguti}}},
  \bibinfo{journal}{\mnras} \textbf{\bibinfo{volume}{318}},
  \bibinfo{pages}{173} (\bibinfo{year}{2000}), \eprint{astro-ph/0005219}.

\bibitem[{\citenamefont{{Marinucci} et~al.}(2016)\citenamefont{{Marinucci},
  {Bianchi}, {Matt}, {Alexander}, {Balokovi{\'c}}, {Bauer}, {Brandt}, {Gandhi},
  {Guainazzi}, {Harrison} et~al.}}]{2016MNRAS.456L..94M}
\bibinfo{author}{\bibfnamefont{A.}~\bibnamefont{{Marinucci}}},
  \bibinfo{author}{\bibfnamefont{S.}~\bibnamefont{{Bianchi}}},
  \bibinfo{author}{\bibfnamefont{G.}~\bibnamefont{{Matt}}},
  \bibinfo{author}{\bibfnamefont{D.~M.} \bibnamefont{{Alexander}}},
  \bibinfo{author}{\bibfnamefont{M.}~\bibnamefont{{Balokovi{\'c}}}},
  \bibinfo{author}{\bibfnamefont{F.~E.} \bibnamefont{{Bauer}}},
  \bibinfo{author}{\bibfnamefont{W.~N.} \bibnamefont{{Brandt}}},
  \bibinfo{author}{\bibfnamefont{P.}~\bibnamefont{{Gandhi}}},
  \bibinfo{author}{\bibfnamefont{M.}~\bibnamefont{{Guainazzi}}},
  \bibinfo{author}{\bibfnamefont{F.~A.} \bibnamefont{{Harrison}}},
  \bibnamefont{et~al.}, \bibinfo{journal}{\mnras}
  \textbf{\bibinfo{volume}{456}}, \bibinfo{pages}{L94} (\bibinfo{year}{2016}),
  \eprint{1511.03503}.

\bibitem[{\citenamefont{{Oh} et~al.}(2018)\citenamefont{{Oh}, {Koss},
  {Markwardt}, {Schawinski}, {Baumgartner}, {Barthelmy}, {Cenko}, {Gehrels},
  {Mushotzky}, {Petulante} et~al.}}]{2018ApJS..235....4O}
\bibinfo{author}{\bibfnamefont{K.}~\bibnamefont{{Oh}}},
  \bibinfo{author}{\bibfnamefont{M.}~\bibnamefont{{Koss}}},
  \bibinfo{author}{\bibfnamefont{C.~B.} \bibnamefont{{Markwardt}}},
  \bibinfo{author}{\bibfnamefont{K.}~\bibnamefont{{Schawinski}}},
  \bibinfo{author}{\bibfnamefont{W.~H.} \bibnamefont{{Baumgartner}}},
  \bibinfo{author}{\bibfnamefont{S.~D.} \bibnamefont{{Barthelmy}}},
  \bibinfo{author}{\bibfnamefont{S.~B.} \bibnamefont{{Cenko}}},
  \bibinfo{author}{\bibfnamefont{N.}~\bibnamefont{{Gehrels}}},
  \bibinfo{author}{\bibfnamefont{R.}~\bibnamefont{{Mushotzky}}},
  \bibinfo{author}{\bibfnamefont{A.}~\bibnamefont{{Petulante}}},
  \bibnamefont{et~al.}, \bibinfo{journal}{\apjs}
  \textbf{\bibinfo{volume}{235}}, \bibinfo{eid}{4} (\bibinfo{year}{2018}),
  \eprint{1801.01882}.

\bibitem[{\citenamefont{Aartsen et~al.}(2020{\natexlab{a}})}]{IceCube:2019cia}
\bibinfo{author}{\bibfnamefont{M.~G.} \bibnamefont{Aartsen}}
  \bibnamefont{et~al.} (\bibinfo{collaboration}{IceCube}),
  \bibinfo{journal}{Phys. Rev. Lett.} \textbf{\bibinfo{volume}{124}},
  \bibinfo{pages}{051103} (\bibinfo{year}{2020}{\natexlab{a}}),
  \eprint{1910.08488}.

\bibitem[{\citenamefont{Aartsen et~al.}(2018)}]{IceCube:2018cha}
\bibinfo{author}{\bibfnamefont{M.~G.} \bibnamefont{Aartsen}}
  \bibnamefont{et~al.} (\bibinfo{collaboration}{IceCube}),
  \bibinfo{journal}{Science} \textbf{\bibinfo{volume}{361}},
  \bibinfo{pages}{147} (\bibinfo{year}{2018}), \eprint{1807.08794}.

\bibitem[{\citenamefont{Abbasi et~al.}(2022)}]{IceCube:2022der}
\bibinfo{author}{\bibfnamefont{R.}~\bibnamefont{Abbasi}} \bibnamefont{et~al.}
  (\bibinfo{collaboration}{IceCube}), \bibinfo{journal}{Science}
  \textbf{\bibinfo{volume}{378}}, \bibinfo{pages}{538} (\bibinfo{year}{2022}),
  \eprint{2211.09972}.

\bibitem[{\citenamefont{{Georgantopoulos} and
  {Akylas}}(2019)}]{2019...621A..28G}
\bibinfo{author}{\bibfnamefont{I.}~\bibnamefont{{Georgantopoulos}}}
  \bibnamefont{and} \bibinfo{author}{\bibfnamefont{A.}~\bibnamefont{{Akylas}}},
  \bibinfo{journal}{A\&A} \textbf{\bibinfo{volume}{621}}, \bibinfo{pages}{A28}
  (\bibinfo{year}{2019}).

\bibitem[{\citenamefont{{Balokovi{\'c}}
  et~al.}(2014)\citenamefont{{Balokovi{\'c}}, {Comastri}, {Harrison},
  {Alexander}, {Ballantyne}, {Bauer}, {Boggs}, {Brandt}, {Brightman},
  {Christensen} et~al.}}]{2014ApJ...794..111B}
\bibinfo{author}{\bibfnamefont{M.}~\bibnamefont{{Balokovi{\'c}}}},
  \bibinfo{author}{\bibfnamefont{A.}~\bibnamefont{{Comastri}}},
  \bibinfo{author}{\bibfnamefont{F.~A.} \bibnamefont{{Harrison}}},
  \bibinfo{author}{\bibfnamefont{D.~M.} \bibnamefont{{Alexander}}},
  \bibinfo{author}{\bibfnamefont{D.~R.} \bibnamefont{{Ballantyne}}},
  \bibinfo{author}{\bibfnamefont{F.~E.} \bibnamefont{{Bauer}}},
  \bibinfo{author}{\bibfnamefont{S.~E.} \bibnamefont{{Boggs}}},
  \bibinfo{author}{\bibfnamefont{W.~N.} \bibnamefont{{Brandt}}},
  \bibinfo{author}{\bibfnamefont{M.}~\bibnamefont{{Brightman}}},
  \bibinfo{author}{\bibfnamefont{F.~E.} \bibnamefont{{Christensen}}},
  \bibnamefont{et~al.}, \bibinfo{journal}{\apj} \textbf{\bibinfo{volume}{794}},
  \bibinfo{eid}{111} (\bibinfo{year}{2014}), \eprint{1408.5414}.

\bibitem[{\citenamefont{{Panagiotou} and {Walter}}(2020)}]{2020...640A..31P}
\bibinfo{author}{\bibfnamefont{C.}~\bibnamefont{{Panagiotou}}}
  \bibnamefont{and} \bibinfo{author}{\bibfnamefont{R.}~\bibnamefont{{Walter}}},
  \bibinfo{journal}{\aap} \textbf{\bibinfo{volume}{640}}, \bibinfo{eid}{A31}
  (\bibinfo{year}{2020}), \eprint{2006.04441}.

\bibitem[{\citenamefont{{Marchesi} et~al.}(2018)\citenamefont{{Marchesi},
  {Ajello}, {Marcotulli}, {Comastri}, {Lanzuisi}, and
  {Vignali}}}]{2018ApJ...854...49M}
\bibinfo{author}{\bibfnamefont{S.}~\bibnamefont{{Marchesi}}},
  \bibinfo{author}{\bibfnamefont{M.}~\bibnamefont{{Ajello}}},
  \bibinfo{author}{\bibfnamefont{L.}~\bibnamefont{{Marcotulli}}},
  \bibinfo{author}{\bibfnamefont{A.}~\bibnamefont{{Comastri}}},
  \bibinfo{author}{\bibfnamefont{G.}~\bibnamefont{{Lanzuisi}}},
  \bibnamefont{and}
  \bibinfo{author}{\bibfnamefont{C.}~\bibnamefont{{Vignali}}},
  \bibinfo{journal}{\apj} \textbf{\bibinfo{volume}{854}}, \bibinfo{eid}{49}
  (\bibinfo{year}{2018}), \eprint{1801.03166}.

\bibitem[{\citenamefont{Bauer et~al.}(2015)}]{Bauer:2014rla}
\bibinfo{author}{\bibfnamefont{F.~E.} \bibnamefont{Bauer}}
  \bibnamefont{et~al.}, \bibinfo{journal}{Astrophys. J.}
  \textbf{\bibinfo{volume}{812}}, \bibinfo{pages}{116} (\bibinfo{year}{2015}),
  \eprint{1411.0670}.

\bibitem[{\citenamefont{{Abbasi} et~al.}(2011)\citenamefont{{Abbasi}, {Abdou},
  {Abu-Zayyad}, {Adams}, {Aguilar}, {Ahlers}, {Andeen}, {Auffenberg}, {Bai},
  {Baker} et~al.}}]{2011PhRvD..83a2001A}
\bibinfo{author}{\bibfnamefont{R.}~\bibnamefont{{Abbasi}}},
  \bibinfo{author}{\bibfnamefont{Y.}~\bibnamefont{{Abdou}}},
  \bibinfo{author}{\bibfnamefont{T.}~\bibnamefont{{Abu-Zayyad}}},
  \bibinfo{author}{\bibfnamefont{J.}~\bibnamefont{{Adams}}},
  \bibinfo{author}{\bibfnamefont{J.~A.} \bibnamefont{{Aguilar}}},
  \bibinfo{author}{\bibfnamefont{M.}~\bibnamefont{{Ahlers}}},
  \bibinfo{author}{\bibfnamefont{K.}~\bibnamefont{{Andeen}}},
  \bibinfo{author}{\bibfnamefont{J.}~\bibnamefont{{Auffenberg}}},
  \bibinfo{author}{\bibfnamefont{X.}~\bibnamefont{{Bai}}},
  \bibinfo{author}{\bibfnamefont{M.}~\bibnamefont{{Baker}}},
  \bibnamefont{et~al.}, \bibinfo{journal}{\prd} \textbf{\bibinfo{volume}{83}},
  \bibinfo{eid}{012001} (\bibinfo{year}{2011}), \eprint{1010.3980}.

\bibitem[{\citenamefont{{Maraschi} and {Haardt}}(1997)}]{1997ASPC..121..101M}
\bibinfo{author}{\bibfnamefont{L.}~\bibnamefont{{Maraschi}}} \bibnamefont{and}
  \bibinfo{author}{\bibfnamefont{F.}~\bibnamefont{{Haardt}}}, in
  \emph{\bibinfo{booktitle}{IAU Colloq. 163: Accretion Phenomena and Related
  Outflows}}, edited by \bibinfo{editor}{\bibfnamefont{D.~T.}
  \bibnamefont{{Wickramasinghe}}}, \bibinfo{editor}{\bibfnamefont{G.~V.}
  \bibnamefont{{Bicknell}}}, \bibnamefont{and}
  \bibinfo{editor}{\bibfnamefont{L.}~\bibnamefont{{Ferrario}}}
  (\bibinfo{year}{1997}), vol. \bibinfo{volume}{121} of
  \emph{\bibinfo{series}{Astronomical Society of the Pacific Conference
  Series}}, p. \bibinfo{pages}{101}, \eprint{astro-ph/9611048}.

\bibitem[{\citenamefont{{Prieto} et~al.}(2010)\citenamefont{{Prieto},
  {Reunanen}, {Tristram}, {Neumayer}, {Fernandez-Ontiveros}, {Orienti}, and
  {Meisenheimer}}}]{2010MNRAS.402..724P}
\bibinfo{author}{\bibfnamefont{M.~A.} \bibnamefont{{Prieto}}},
  \bibinfo{author}{\bibfnamefont{J.}~\bibnamefont{{Reunanen}}},
  \bibinfo{author}{\bibfnamefont{K.~R.~W.} \bibnamefont{{Tristram}}},
  \bibinfo{author}{\bibfnamefont{N.}~\bibnamefont{{Neumayer}}},
  \bibinfo{author}{\bibfnamefont{J.~A.} \bibnamefont{{Fernandez-Ontiveros}}},
  \bibinfo{author}{\bibfnamefont{M.}~\bibnamefont{{Orienti}}},
  \bibnamefont{and}
  \bibinfo{author}{\bibfnamefont{K.}~\bibnamefont{{Meisenheimer}}},
  \bibinfo{journal}{\mnras} \textbf{\bibinfo{volume}{402}},
  \bibinfo{pages}{724} (\bibinfo{year}{2010}), \eprint{0910.3771}.

\bibitem[{\citenamefont{{Balokovic}}(2017)}]{2017PhDT........40B}
\bibinfo{author}{\bibfnamefont{M.}~\bibnamefont{{Balokovic}}}, Ph.D. thesis,
  \bibinfo{school}{California Institute of Technology} (\bibinfo{year}{2017}).

\bibitem[{\citenamefont{{Pizzetti} et~al.}(2022)\citenamefont{{Pizzetti},
  {Torres-Alb{\`a}}, {Marchesi}, {Ajello}, {Silver}, and
  {Zhao}}}]{2022ApJ...936..149P}
\bibinfo{author}{\bibfnamefont{A.}~\bibnamefont{{Pizzetti}}},
  \bibinfo{author}{\bibfnamefont{N.}~\bibnamefont{{Torres-Alb{\`a}}}},
  \bibinfo{author}{\bibfnamefont{S.}~\bibnamefont{{Marchesi}}},
  \bibinfo{author}{\bibfnamefont{M.}~\bibnamefont{{Ajello}}},
  \bibinfo{author}{\bibfnamefont{R.}~\bibnamefont{{Silver}}}, \bibnamefont{and}
  \bibinfo{author}{\bibfnamefont{X.}~\bibnamefont{{Zhao}}},
  \bibinfo{journal}{\apj} \textbf{\bibinfo{volume}{936}}, \bibinfo{eid}{149}
  (\bibinfo{year}{2022}), \eprint{2206.10946}.

\bibitem[{\citenamefont{{IceCube Collaboration}
  et~al.}(2021)\citenamefont{{IceCube Collaboration}, {Abbasi}, {Ackermann},
  {Adams}, {Aguilar}, {Ahlers}, {Ahrens}, {Alispach}, {Amin}, {Andeen}
  et~al.}}]{2021arXiv210109836I}
\bibinfo{author}{\bibnamefont{{IceCube Collaboration}}},
  \bibinfo{author}{\bibfnamefont{R.}~\bibnamefont{{Abbasi}}},
  \bibinfo{author}{\bibfnamefont{M.}~\bibnamefont{{Ackermann}}},
  \bibinfo{author}{\bibfnamefont{J.}~\bibnamefont{{Adams}}},
  \bibinfo{author}{\bibfnamefont{J.~A.} \bibnamefont{{Aguilar}}},
  \bibinfo{author}{\bibfnamefont{M.}~\bibnamefont{{Ahlers}}},
  \bibinfo{author}{\bibfnamefont{M.}~\bibnamefont{{Ahrens}}},
  \bibinfo{author}{\bibfnamefont{C.}~\bibnamefont{{Alispach}}},
  \bibinfo{author}{\bibfnamefont{N.~M.} \bibnamefont{{Amin}}},
  \bibinfo{author}{\bibfnamefont{K.}~\bibnamefont{{Andeen}}},
  \bibnamefont{et~al.}, \bibinfo{journal}{arXiv e-prints}
  \bibinfo{eid}{arXiv:2101.09836} (\bibinfo{year}{2021}), \eprint{2101.09836}.

\bibitem[{\citenamefont{{Mattox} et~al.}(1996)\citenamefont{{Mattox},
  {Bertsch}, {Chiang}, {Dingus}, {Digel}, {Esposito}, {Fierro}, {Hartman},
  {Hunter}, {Kanbach} et~al.}}]{mattox96}
\bibinfo{author}{\bibfnamefont{J.~R.} \bibnamefont{{Mattox}}},
  \bibinfo{author}{\bibfnamefont{D.~L.} \bibnamefont{{Bertsch}}},
  \bibinfo{author}{\bibfnamefont{J.}~\bibnamefont{{Chiang}}},
  \bibinfo{author}{\bibfnamefont{B.~L.} \bibnamefont{{Dingus}}},
  \bibinfo{author}{\bibfnamefont{S.~W.} \bibnamefont{{Digel}}},
  \bibinfo{author}{\bibfnamefont{J.~A.} \bibnamefont{{Esposito}}},
  \bibinfo{author}{\bibfnamefont{J.~M.} \bibnamefont{{Fierro}}},
  \bibinfo{author}{\bibfnamefont{R.~C.} \bibnamefont{{Hartman}}},
  \bibinfo{author}{\bibfnamefont{S.~D.} \bibnamefont{{Hunter}}},
  \bibinfo{author}{\bibfnamefont{G.}~\bibnamefont{{Kanbach}}},
  \bibnamefont{et~al.}, \bibinfo{journal}{\apj} \textbf{\bibinfo{volume}{461}},
  \bibinfo{pages}{396} (\bibinfo{year}{1996}).

\bibitem[{\citenamefont{Marinucci et~al.}(2018)\citenamefont{Marinucci,
  Tamborra, Bianchi, Dovčiak, Matt, Middei, and Tortosa}}]{galaxies6020044}
\bibinfo{author}{\bibfnamefont{A.}~\bibnamefont{Marinucci}},
  \bibinfo{author}{\bibfnamefont{F.}~\bibnamefont{Tamborra}},
  \bibinfo{author}{\bibfnamefont{S.}~\bibnamefont{Bianchi}},
  \bibinfo{author}{\bibfnamefont{M.}~\bibnamefont{Dovčiak}},
  \bibinfo{author}{\bibfnamefont{G.}~\bibnamefont{Matt}},
  \bibinfo{author}{\bibfnamefont{R.}~\bibnamefont{Middei}}, \bibnamefont{and}
  \bibinfo{author}{\bibfnamefont{A.}~\bibnamefont{Tortosa}},
  \bibinfo{journal}{Galaxies} \textbf{\bibinfo{volume}{6}}
  (\bibinfo{year}{2018}), ISSN \bibinfo{issn}{2075-4434},
  \urlprefix\url{https://www.mdpi.com/2075-4434/6/2/44}.

\bibitem[{\citenamefont{Churazov et~al.}(2007)}]{Churazov:2006bk}
\bibinfo{author}{\bibfnamefont{E.}~\bibnamefont{Churazov}}
  \bibnamefont{et~al.}, \bibinfo{journal}{Astron. Astrophys.}
  \textbf{\bibinfo{volume}{467}}, \bibinfo{pages}{529} (\bibinfo{year}{2007}),
  \eprint{astro-ph/0608250}.

\bibitem[{\citenamefont{Aartsen et~al.}(2020{\natexlab{b}})}]{IceCube:2020acn}
\bibinfo{author}{\bibfnamefont{M.~G.} \bibnamefont{Aartsen}}
  \bibnamefont{et~al.} (\bibinfo{collaboration}{IceCube}),
  \bibinfo{journal}{Phys. Rev. Lett.} \textbf{\bibinfo{volume}{125}},
  \bibinfo{pages}{121104} (\bibinfo{year}{2020}{\natexlab{b}}),
  \eprint{2001.09520}.

\end{thebibliography}


\begin{thebibliography}{18}
\expandafter\ifx\csname natexlab\endcsname\relax\def\natexlab#1{#1}\fi
\expandafter\ifx\csname bibnamefont\endcsname\relax
  \def\bibnamefont#1{#1}\fi
\expandafter\ifx\csname bibfnamefont\endcsname\relax
  \def\bibfnamefont#1{#1}\fi
\expandafter\ifx\csname citenamefont\endcsname\relax
  \def\citenamefont#1{#1}\fi
\expandafter\ifx\csname url\endcsname\relax
  \def\url#1{\texttt{#1}}\fi
\expandafter\ifx\csname urlprefix\endcsname\relax\def\urlprefix{URL }\fi
\providecommand{\bibinfo}[2]{#2}
\providecommand{\eprint}[2][]{\url{#2}}

\bibitem[{\citenamefont{{Fischer} et~al.}(2021)\citenamefont{{Fischer},
  {Secrest}, {Johnson}, {Dorland}, {Cigan}, {Fernandez}, {Hunt}, {Koss},
  {Schmitt}, and {Zacharias}}}]{2021ApJ...906...88F}
\bibinfo{author}{\bibfnamefont{T.~C.} \bibnamefont{{Fischer}}},
  \bibinfo{author}{\bibfnamefont{N.~J.} \bibnamefont{{Secrest}}},
  \bibinfo{author}{\bibfnamefont{M.~C.} \bibnamefont{{Johnson}}},
  \bibinfo{author}{\bibfnamefont{B.~N.} \bibnamefont{{Dorland}}},
  \bibinfo{author}{\bibfnamefont{P.~J.} \bibnamefont{{Cigan}}},
  \bibinfo{author}{\bibfnamefont{L.~C.} \bibnamefont{{Fernandez}}},
  \bibinfo{author}{\bibfnamefont{L.~R.} \bibnamefont{{Hunt}}},
  \bibinfo{author}{\bibfnamefont{M.}~\bibnamefont{{Koss}}},
  \bibinfo{author}{\bibfnamefont{H.~R.} \bibnamefont{{Schmitt}}},
  \bibnamefont{and}
  \bibinfo{author}{\bibfnamefont{N.}~\bibnamefont{{Zacharias}}},
  \bibinfo{journal}{\apj} \textbf{\bibinfo{volume}{906}}, \bibinfo{eid}{88}
  (\bibinfo{year}{2021}), \eprint{2011.06570}.

\bibitem[{\citenamefont{{Pe{\~n}a-Herazo}
  et~al.}(2022)\citenamefont{{Pe{\~n}a-Herazo}, {Massaro}, {Chavushyan},
  {Masetti}, {Paggi}, and {Capetti}}}]{2022A&A...659A..32P}
\bibinfo{author}{\bibfnamefont{H.~A.} \bibnamefont{{Pe{\~n}a-Herazo}}},
  \bibinfo{author}{\bibfnamefont{F.}~\bibnamefont{{Massaro}}},
  \bibinfo{author}{\bibfnamefont{V.}~\bibnamefont{{Chavushyan}}},
  \bibinfo{author}{\bibfnamefont{N.}~\bibnamefont{{Masetti}}},
  \bibinfo{author}{\bibfnamefont{A.}~\bibnamefont{{Paggi}}}, \bibnamefont{and}
  \bibinfo{author}{\bibfnamefont{A.}~\bibnamefont{{Capetti}}},
  \bibinfo{journal}{\aap} \textbf{\bibinfo{volume}{659}}, \bibinfo{eid}{A32}
  (\bibinfo{year}{2022}).

\bibitem[{\citenamefont{{Marinucci} et~al.}(2016)\citenamefont{{Marinucci},
  {Bianchi}, {Matt}, {Alexander}, {Balokovi{\'c}}, {Bauer}, {Brandt}, {Gandhi},
  {Guainazzi}, {Harrison} et~al.}}]{2016MNRAS.456L..94M}
\bibinfo{author}{\bibfnamefont{A.}~\bibnamefont{{Marinucci}}},
  \bibinfo{author}{\bibfnamefont{S.}~\bibnamefont{{Bianchi}}},
  \bibinfo{author}{\bibfnamefont{G.}~\bibnamefont{{Matt}}},
  \bibinfo{author}{\bibfnamefont{D.~M.} \bibnamefont{{Alexander}}},
  \bibinfo{author}{\bibfnamefont{M.}~\bibnamefont{{Balokovi{\'c}}}},
  \bibinfo{author}{\bibfnamefont{F.~E.} \bibnamefont{{Bauer}}},
  \bibinfo{author}{\bibfnamefont{W.~N.} \bibnamefont{{Brandt}}},
  \bibinfo{author}{\bibfnamefont{P.}~\bibnamefont{{Gandhi}}},
  \bibinfo{author}{\bibfnamefont{M.}~\bibnamefont{{Guainazzi}}},
  \bibinfo{author}{\bibfnamefont{F.~A.} \bibnamefont{{Harrison}}},
  \bibnamefont{et~al.}, \bibinfo{journal}{\mnras}
  \textbf{\bibinfo{volume}{456}}, \bibinfo{pages}{L94} (\bibinfo{year}{2016}),
  \eprint{1511.03503}.

\bibitem[{\citenamefont{{Georgantopoulos} and
  {Akylas}}(2019)}]{2019...621A..28G}
\bibinfo{author}{\bibfnamefont{I.}~\bibnamefont{{Georgantopoulos}}}
  \bibnamefont{and} \bibinfo{author}{\bibfnamefont{A.}~\bibnamefont{{Akylas}}},
  \bibinfo{journal}{A\&A} \textbf{\bibinfo{volume}{621}}, \bibinfo{pages}{A28}
  (\bibinfo{year}{2019}).

\bibitem[{\citenamefont{{Balokovic}}(2017)}]{2017PhDT........40B}
\bibinfo{author}{\bibfnamefont{M.}~\bibnamefont{{Balokovic}}}, Ph.D. thesis,
  \bibinfo{school}{California Institute of Technology} (\bibinfo{year}{2017}).

\bibitem[{\citenamefont{{Balokovi{\'c}}
  et~al.}(2014)\citenamefont{{Balokovi{\'c}}, {Comastri}, {Harrison},
  {Alexander}, {Ballantyne}, {Bauer}, {Boggs}, {Brandt}, {Brightman},
  {Christensen} et~al.}}]{2014ApJ...794..111B}
\bibinfo{author}{\bibfnamefont{M.}~\bibnamefont{{Balokovi{\'c}}}},
  \bibinfo{author}{\bibfnamefont{A.}~\bibnamefont{{Comastri}}},
  \bibinfo{author}{\bibfnamefont{F.~A.} \bibnamefont{{Harrison}}},
  \bibinfo{author}{\bibfnamefont{D.~M.} \bibnamefont{{Alexander}}},
  \bibinfo{author}{\bibfnamefont{D.~R.} \bibnamefont{{Ballantyne}}},
  \bibinfo{author}{\bibfnamefont{F.~E.} \bibnamefont{{Bauer}}},
  \bibinfo{author}{\bibfnamefont{S.~E.} \bibnamefont{{Boggs}}},
  \bibinfo{author}{\bibfnamefont{W.~N.} \bibnamefont{{Brandt}}},
  \bibinfo{author}{\bibfnamefont{M.}~\bibnamefont{{Brightman}}},
  \bibinfo{author}{\bibfnamefont{F.~E.} \bibnamefont{{Christensen}}},
  \bibnamefont{et~al.}, \bibinfo{journal}{\apj} \textbf{\bibinfo{volume}{794}},
  \bibinfo{eid}{111} (\bibinfo{year}{2014}), \eprint{1408.5414}.

\bibitem[{\citenamefont{{Panagiotou} and {Walter}}(2020)}]{2020...640A..31P}
\bibinfo{author}{\bibfnamefont{C.}~\bibnamefont{{Panagiotou}}}
  \bibnamefont{and} \bibinfo{author}{\bibfnamefont{R.}~\bibnamefont{{Walter}}},
  \bibinfo{journal}{\aap} \textbf{\bibinfo{volume}{640}}, \bibinfo{eid}{A31}
  (\bibinfo{year}{2020}), \eprint{2006.04441}.

\bibitem[{\citenamefont{{Marchesi} et~al.}(2018)\citenamefont{{Marchesi},
  {Ajello}, {Marcotulli}, {Comastri}, {Lanzuisi}, and
  {Vignali}}}]{2018ApJ...854...49M}
\bibinfo{author}{\bibfnamefont{S.}~\bibnamefont{{Marchesi}}},
  \bibinfo{author}{\bibfnamefont{M.}~\bibnamefont{{Ajello}}},
  \bibinfo{author}{\bibfnamefont{L.}~\bibnamefont{{Marcotulli}}},
  \bibinfo{author}{\bibfnamefont{A.}~\bibnamefont{{Comastri}}},
  \bibinfo{author}{\bibfnamefont{G.}~\bibnamefont{{Lanzuisi}}},
  \bibnamefont{and}
  \bibinfo{author}{\bibfnamefont{C.}~\bibnamefont{{Vignali}}},
  \bibinfo{journal}{\apj} \textbf{\bibinfo{volume}{854}}, \bibinfo{eid}{49}
  (\bibinfo{year}{2018}), \eprint{1801.03166}.

\bibitem[{\citenamefont{{Pizzetti} et~al.}(2022)\citenamefont{{Pizzetti},
  {Torres-Alb{\`a}}, {Marchesi}, {Ajello}, {Silver}, and
  {Zhao}}}]{2022ApJ...936..149P}
\bibinfo{author}{\bibfnamefont{A.}~\bibnamefont{{Pizzetti}}},
  \bibinfo{author}{\bibfnamefont{N.}~\bibnamefont{{Torres-Alb{\`a}}}},
  \bibinfo{author}{\bibfnamefont{S.}~\bibnamefont{{Marchesi}}},
  \bibinfo{author}{\bibfnamefont{M.}~\bibnamefont{{Ajello}}},
  \bibinfo{author}{\bibfnamefont{R.}~\bibnamefont{{Silver}}}, \bibnamefont{and}
  \bibinfo{author}{\bibfnamefont{X.}~\bibnamefont{{Zhao}}},
  \bibinfo{journal}{\apj} \textbf{\bibinfo{volume}{936}}, \bibinfo{eid}{149}
  (\bibinfo{year}{2022}), \eprint{2206.10946}.

\bibitem[{\citenamefont{{Oh} et~al.}(2018)\citenamefont{{Oh}, {Koss},
  {Markwardt}, {Schawinski}, {Baumgartner}, {Barthelmy}, {Cenko}, {Gehrels},
  {Mushotzky}, {Petulante} et~al.}}]{2018ApJS..235....4O}
\bibinfo{author}{\bibfnamefont{K.}~\bibnamefont{{Oh}}},
  \bibinfo{author}{\bibfnamefont{M.}~\bibnamefont{{Koss}}},
  \bibinfo{author}{\bibfnamefont{C.~B.} \bibnamefont{{Markwardt}}},
  \bibinfo{author}{\bibfnamefont{K.}~\bibnamefont{{Schawinski}}},
  \bibinfo{author}{\bibfnamefont{W.~H.} \bibnamefont{{Baumgartner}}},
  \bibinfo{author}{\bibfnamefont{S.~D.} \bibnamefont{{Barthelmy}}},
  \bibinfo{author}{\bibfnamefont{S.~B.} \bibnamefont{{Cenko}}},
  \bibinfo{author}{\bibfnamefont{N.}~\bibnamefont{{Gehrels}}},
  \bibinfo{author}{\bibfnamefont{R.}~\bibnamefont{{Mushotzky}}},
  \bibinfo{author}{\bibfnamefont{A.}~\bibnamefont{{Petulante}}},
  \bibnamefont{et~al.}, \bibinfo{journal}{\apjs}
  \textbf{\bibinfo{volume}{235}}, \bibinfo{eid}{4} (\bibinfo{year}{2018}),
  \eprint{1801.01882}.

\bibitem[{\citenamefont{Bauer et~al.}(2015)}]{Bauer:2014rla}
\bibinfo{author}{\bibfnamefont{F.~E.} \bibnamefont{Bauer}}
  \bibnamefont{et~al.}, \bibinfo{journal}{Astrophys. J.}
  \textbf{\bibinfo{volume}{812}}, \bibinfo{pages}{116} (\bibinfo{year}{2015}),
  \eprint{1411.0670}.

\bibitem[{\citenamefont{{Mattox} et~al.}(1996)\citenamefont{{Mattox},
  {Bertsch}, {Chiang}, {Dingus}, {Digel}, {Esposito}, {Fierro}, {Hartman},
  {Hunter}, {Kanbach} et~al.}}]{mattox96}
\bibinfo{author}{\bibfnamefont{J.~R.} \bibnamefont{{Mattox}}},
  \bibinfo{author}{\bibfnamefont{D.~L.} \bibnamefont{{Bertsch}}},
  \bibinfo{author}{\bibfnamefont{J.}~\bibnamefont{{Chiang}}},
  \bibinfo{author}{\bibfnamefont{B.~L.} \bibnamefont{{Dingus}}},
  \bibinfo{author}{\bibfnamefont{S.~W.} \bibnamefont{{Digel}}},
  \bibinfo{author}{\bibfnamefont{J.~A.} \bibnamefont{{Esposito}}},
  \bibinfo{author}{\bibfnamefont{J.~M.} \bibnamefont{{Fierro}}},
  \bibinfo{author}{\bibfnamefont{R.~C.} \bibnamefont{{Hartman}}},
  \bibinfo{author}{\bibfnamefont{S.~D.} \bibnamefont{{Hunter}}},
  \bibinfo{author}{\bibfnamefont{G.}~\bibnamefont{{Kanbach}}},
  \bibnamefont{et~al.}, \bibinfo{journal}{\apj} \textbf{\bibinfo{volume}{461}},
  \bibinfo{pages}{396} (\bibinfo{year}{1996}).

\bibitem[{\citenamefont{{Abbasi} et~al.}(2011)\citenamefont{{Abbasi}, {Abdou},
  {Abu-Zayyad}, {Adams}, {Aguilar}, {Ahlers}, {Andeen}, {Auffenberg}, {Bai},
  {Baker} et~al.}}]{2011PhRvD..83a2001A}
\bibinfo{author}{\bibfnamefont{R.}~\bibnamefont{{Abbasi}}},
  \bibinfo{author}{\bibfnamefont{Y.}~\bibnamefont{{Abdou}}},
  \bibinfo{author}{\bibfnamefont{T.}~\bibnamefont{{Abu-Zayyad}}},
  \bibinfo{author}{\bibfnamefont{J.}~\bibnamefont{{Adams}}},
  \bibinfo{author}{\bibfnamefont{J.~A.} \bibnamefont{{Aguilar}}},
  \bibinfo{author}{\bibfnamefont{M.}~\bibnamefont{{Ahlers}}},
  \bibinfo{author}{\bibfnamefont{K.}~\bibnamefont{{Andeen}}},
  \bibinfo{author}{\bibfnamefont{J.}~\bibnamefont{{Auffenberg}}},
  \bibinfo{author}{\bibfnamefont{X.}~\bibnamefont{{Bai}}},
  \bibinfo{author}{\bibfnamefont{M.}~\bibnamefont{{Baker}}},
  \bibnamefont{et~al.}, \bibinfo{journal}{\prd} \textbf{\bibinfo{volume}{83}},
  \bibinfo{eid}{012001} (\bibinfo{year}{2011}), \eprint{1010.3980}.

\bibitem[{\citenamefont{{Aartsen} et~al.}(2020)\citenamefont{{Aartsen},
  {Ackermann}, {Adams}, {Aguilar}, {Ahlers}, {Ahrens}, {Alispach}, {Andeen},
  {Anderson}, {Ansseau} et~al.}}]{icecube_10yr_paper}
\bibinfo{author}{\bibfnamefont{M.~G.} \bibnamefont{{Aartsen}}},
  \bibinfo{author}{\bibfnamefont{M.}~\bibnamefont{{Ackermann}}},
  \bibinfo{author}{\bibfnamefont{J.}~\bibnamefont{{Adams}}},
  \bibinfo{author}{\bibfnamefont{J.~A.} \bibnamefont{{Aguilar}}},
  \bibinfo{author}{\bibfnamefont{M.}~\bibnamefont{{Ahlers}}},
  \bibinfo{author}{\bibfnamefont{M.}~\bibnamefont{{Ahrens}}},
  \bibinfo{author}{\bibfnamefont{C.}~\bibnamefont{{Alispach}}},
  \bibinfo{author}{\bibfnamefont{K.}~\bibnamefont{{Andeen}}},
  \bibinfo{author}{\bibfnamefont{T.}~\bibnamefont{{Anderson}}},
  \bibinfo{author}{\bibfnamefont{I.}~\bibnamefont{{Ansseau}}},
  \bibnamefont{et~al.}, \bibinfo{journal}{\prl} \textbf{\bibinfo{volume}{124}},
  \bibinfo{eid}{051103} (\bibinfo{year}{2020}), \eprint{1910.08488}.

\bibitem[{\citenamefont{{IceCube Collaboration}
  et~al.}(2021)\citenamefont{{IceCube Collaboration}, {Abbasi}, {Ackermann},
  {Adams}, {Aguilar}, {Ahlers}, {Ahrens}, {Alispach}, {Amin}, {Andeen}
  et~al.}}]{2021arXiv210109836I}
\bibinfo{author}{\bibnamefont{{IceCube Collaboration}}},
  \bibinfo{author}{\bibfnamefont{R.}~\bibnamefont{{Abbasi}}},
  \bibinfo{author}{\bibfnamefont{M.}~\bibnamefont{{Ackermann}}},
  \bibinfo{author}{\bibfnamefont{J.}~\bibnamefont{{Adams}}},
  \bibinfo{author}{\bibfnamefont{J.~A.} \bibnamefont{{Aguilar}}},
  \bibinfo{author}{\bibfnamefont{M.}~\bibnamefont{{Ahlers}}},
  \bibinfo{author}{\bibfnamefont{M.}~\bibnamefont{{Ahrens}}},
  \bibinfo{author}{\bibfnamefont{C.}~\bibnamefont{{Alispach}}},
  \bibinfo{author}{\bibfnamefont{N.~M.} \bibnamefont{{Amin}}},
  \bibinfo{author}{\bibfnamefont{K.}~\bibnamefont{{Andeen}}},
  \bibnamefont{et~al.}, \bibinfo{journal}{arXiv e-prints}
  \bibinfo{eid}{arXiv:2101.09836} (\bibinfo{year}{2021}), \eprint{2101.09836}.

\bibitem[{\citenamefont{{IceCube Collaboration}
  et~al.}(2022)\citenamefont{{IceCube Collaboration}, {Abbasi}, {Ackermann},
  {Adams}, {Aguilar}, {Ahlers}, {Ahrens}, {Alameddine}, {Alispach}, {Alves}
  et~al.}}]{2022Sci...378..538I}
\bibinfo{author}{\bibnamefont{{IceCube Collaboration}}},
  \bibinfo{author}{\bibfnamefont{R.}~\bibnamefont{{Abbasi}}},
  \bibinfo{author}{\bibfnamefont{M.}~\bibnamefont{{Ackermann}}},
  \bibinfo{author}{\bibfnamefont{J.}~\bibnamefont{{Adams}}},
  \bibinfo{author}{\bibfnamefont{J.~A.} \bibnamefont{{Aguilar}}},
  \bibinfo{author}{\bibfnamefont{M.}~\bibnamefont{{Ahlers}}},
  \bibinfo{author}{\bibfnamefont{M.}~\bibnamefont{{Ahrens}}},
  \bibinfo{author}{\bibfnamefont{J.~M.} \bibnamefont{{Alameddine}}},
  \bibinfo{author}{\bibfnamefont{C.}~\bibnamefont{{Alispach}}},
  \bibinfo{author}{\bibfnamefont{J.}~\bibnamefont{{Alves}},
  \bibfnamefont{A.~A.}}, \bibnamefont{et~al.}, \bibinfo{journal}{Science}
  \textbf{\bibinfo{volume}{378}}, \bibinfo{pages}{538} (\bibinfo{year}{2022}),
  \eprint{2211.09972}.

\bibitem[{\citenamefont{{Maraschi} and {Haardt}}(1997)}]{1997ASPC..121..101M}
\bibinfo{author}{\bibfnamefont{L.}~\bibnamefont{{Maraschi}}} \bibnamefont{and}
  \bibinfo{author}{\bibfnamefont{F.}~\bibnamefont{{Haardt}}}, in
  \emph{\bibinfo{booktitle}{IAU Colloq. 163: Accretion Phenomena and Related
  Outflows}}, edited by \bibinfo{editor}{\bibfnamefont{D.~T.}
  \bibnamefont{{Wickramasinghe}}}, \bibinfo{editor}{\bibfnamefont{G.~V.}
  \bibnamefont{{Bicknell}}}, \bibnamefont{and}
  \bibinfo{editor}{\bibfnamefont{L.}~\bibnamefont{{Ferrario}}}
  (\bibinfo{year}{1997}), vol. \bibinfo{volume}{121} of
  \emph{\bibinfo{series}{Astronomical Society of the Pacific Conference
  Series}}, p. \bibinfo{pages}{101}, \eprint{astro-ph/9611048}.

\bibitem[{\citenamefont{{Prieto} et~al.}(2010)\citenamefont{{Prieto},
  {Reunanen}, {Tristram}, {Neumayer}, {Fernandez-Ontiveros}, {Orienti}, and
  {Meisenheimer}}}]{2010MNRAS.402..724P}
\bibinfo{author}{\bibfnamefont{M.~A.} \bibnamefont{{Prieto}}},
  \bibinfo{author}{\bibfnamefont{J.}~\bibnamefont{{Reunanen}}},
  \bibinfo{author}{\bibfnamefont{K.~R.~W.} \bibnamefont{{Tristram}}},
  \bibinfo{author}{\bibfnamefont{N.}~\bibnamefont{{Neumayer}}},
  \bibinfo{author}{\bibfnamefont{J.~A.} \bibnamefont{{Fernandez-Ontiveros}}},
  \bibinfo{author}{\bibfnamefont{M.}~\bibnamefont{{Orienti}}},
  \bibnamefont{and}
  \bibinfo{author}{\bibfnamefont{K.}~\bibnamefont{{Meisenheimer}}},
  \bibinfo{journal}{\mnras} \textbf{\bibinfo{volume}{402}},
  \bibinfo{pages}{724} (\bibinfo{year}{2010}), \eprint{0910.3771}.

\end{thebibliography}

\end{document}


\title{Supplementary material}
\author{A.~Neronov$^{1,2}$}
\author{D.~Savchenko$^{1,3,4}$}
\author{D.~V.~Semikoz$^{1}$}

\affiliation{$^{1}$Universit\'e Paris Cit\'e, CNRS, Astroparticule et Cosmologie, 
F-75013 Paris, France}

\affiliation{$^{2}$Laboratory of Astrophysics, \'Ecole Polytechnique F\'ed\'erale de Lausanne, CH-1015 Lausanne, Switzerland}

\affiliation{$^{3}$Bogolyubov Institute for Theoretical Physics of the NAS of Ukraine, 03143 Kyiv, Ukraine}

\affiliation{$^{4}$Kyiv Academic University, 03142 Kyiv, Ukraine}
\maketitle

\section{Source list}

Table \ref{tab:list} presents the list of sources considered in the analysis. As described in the main text, the source selection is based on Ref. \cite{2021ApJ...906...88F}  that has presented a volume complete sample of Seyfert galaxies with luminosities above $10^{42}$~erg/s within the declination band $-30^\circ<\delta<60^\circ$. We have removed from this source sample sources at declinations $\delta<-5^\circ$ that cannot be observed with sufficiently low energy threshold in the muon detection channel of IceCube. We have also removed one source at declination $\delta>-5^\circ$, NGC 4765, because it is not listed as a Seyfert galaxy in the catalog of Ref. \cite{2022A&A...659A..32P} and in SIMBAD database (https://simbad.unistra.fr/simbad/). 

\begin{table*}
\begin{tabular}{lllllllll}
\hline
Name & $RA$ &$Dec$ & $D$ & $F_{hX}$ &$L_{hX0}$ & $N_H$&Type \\
 &  &  & &14-195 keV & 14-195 keV & \\
&  & &Mpc & $10^{-11}\frac{\mbox{erg}}{\mbox{cm}^2\mbox{s}}$ & $10^{43}\frac{\mbox{erg}}{\mbox{s}}$ &$10^{24}$~cm$^2$ \\
\hline
NGC 1068 &  $40.6696342$ & $-0.01323785$ & $16.3$ &  $3.79$& $5-22$ \cite{2016MNRAS.456L..94M} & $>10$ 
\cite{2019...621A..28G} 
& Sy2\\
NGC 1320& $51.2028681$ & $-3.04226840$ &$38.4$ & $1.31$ &$0.27$ \cite{2017PhDT........40B}\footnote{Recalculated to 14-195~keV assuming $E^{-2}$ spectrum.}&$3-6$ \cite{2014ApJ...794..111B}&Sy2\\
IC 2461 &  $139.9914308$ & $+37.19100007$& $32.3$ & $1.91$&&$0.08$ \cite{2020...640A..31P}&Sy2\\
NGC 3079 &  $150.4908469$ & $+55.67979744$ & $15.9$ & $3.67$ &$1.0-1.6$ \cite{2017PhDT........40B}  &$2.5$ \cite{2018ApJ...854...49M}, $3.2$ \cite{2019...621A..28G}, $8.5$\cite{2021ApJ...906...88F} &Sy2\\
NGC 3227 & $155.8774015$ & $+19.86505766$ & $16.8$ & 11.24 && $0.009-0.07$ \cite{2020...640A..31P} &Sy1\\
NGC 3786 &  $174.9271391$ & $+31.90942732$ & $38.4$ & $1.46$ &&$0.02$ \cite{2020...640A..31P}&Sy2\\
NGC 4151 &  $182.6357547$ & $+39.40584860$ & 14.2 & $61.89$& &$0.08$ \cite{2018ApJ...854...49M}&Sy1\\
NGC 4235 &  $184.2911678$ & $+7.19157597$ & $34.5$ & $3.86$& & $0.003$ \cite{2018ApJ...854...49M}& Sy1 \\
NGC 4388 &  $186.4449188$ & $+12.66215153$ & $36.2$ & $27.89$ &$1.4-1.5$ \cite{2017PhDT........40B} &$0.5$ \cite{2018ApJ...854...49M} &Sy2 \\
NGC 5290 &  $206.3297085$ & $+41.71241871$ & $37.1$ & $1.49$ && $0.0095$ \cite{2018ApJ...854...49M}& Sy2\\
NGC 5506 & $213.3119888$ & $-3.20768334$ & $26.7$ & $23.94$& &$0.012$ \cite{2018ApJ...854...49M}&Sy1.9\\
NGC 5899 &  $228.7634964$ & $+42.04991289$ & $37.1$ &  $2.04$&$0.3$ \cite{2017PhDT........40B} &$0.11$ \cite{2018ApJ...854...49M}&Sy2\\
NGC 7479 &  $346.2359605$ & $+12.32295297$ &  $34.0$ & $1.69$& $0.9$ \cite{2022ApJ...936..149P}& $5.7$ \cite{2018ApJ...854...49M} &Sy2\\
\hline
\end{tabular}
\caption{ Volume complete sample of Seyfert galaxies with luminosity $L_{hX}>10^{42}$~erg/s in $-5^\circ<\delta<60^\circ$ declination strip, from Ref. \cite{2021ApJ...906...88F}. $Ra, Dec$, distances $D$ and Seyfert types  are from \cite{2021ApJ...906...88F}. 14-195~keV fluxes $F_{hX}$ are from \cite{2018ApJS..235....4O}. 
}
\label{tab:list}
\end{table*}

As discussed in the main text, hard X-ray flux from sources with large hydrogen column density may be affected by Compton scattering that attenuates the direct flux from the primary source and adds an additional reflected flux component. This complicates estimation of the power of the primary hard X-ray source (presumably a hot corona around the black hole). For sources with high $N_H>10^{23}$~cm$^2$, Table \ref{tab:list} provides estimates of the luminosity of the hot corona  found in the literature reporting analysis of NUSTAR data. 

For NGC 1068, we use Ref. \cite{2016MNRAS.456L..94M} that provides an estimate of the intrinsic luminosity in the 2-10~keV band. We re-calculate this luminosity into the Swift/BAT energy band 15-195~keV by assuming that the intrinsic spectrum is an $E^{-2}$ powerlaw, a model consistent with  the spectral modelling of Ref. \cite{Bauer:2014rla} that was used in Ref.  \cite{2016MNRAS.456L..94M}.  For NGC 1320, NGC 3079, NGC 4388 and NGC 5899, we rely on analysis of  NuSTAR data reported in Ref. \cite{2017PhDT........40B}. This reference presents the intrinsic luminosity in the energy range 10-50~keV, which we also re-calculate to the 15-195 keV band assuming an $E^{-2}$ powerlaw spectrum. This spectrum is also consistent with the spectral modelling of Ref.  \cite{2017PhDT........40B} in which the intrinsic spectra of the hot corona were assumed to have no cut-off.  For NGC 7479 we rely on the modelling of Ref.  \cite{2022ApJ...936..149P}  for which we also re-cast the 10-40~keV luminosity estimate to 15-195~keV range assuming an $E^{-2}$ powerlaw for the intrinsic source spectrum.

\section{IceCube data analysis}

We use the 10-year publicly available IceCube data set and perform the unbinned likelihood analysis of the data \cite{mattox96}. We calculate the likelihood function
\begin{equation}
\label{eq:likelihood}
\log L(N_s)=\sum_i\log\left(\frac{N_s}{N_t}S_i+\left(1-\frac{N_s}{N_t}\right)B_i\right)
\end{equation}
where the sum runs over the neutrino event sample,  $N_s,N_t$ are the source and total event counts, \linebreak $S_i$ and $B_i$ are the  probability density functions  (PDF) of the signal and background for $i$-th event. The sum runs through all the events in a Region Of Interest (ROI) which we choose to be a $10^\circ$ large square aligned along $RA,Dec$ directions, centered on the source. 

The background depends on the declination, because of the geographical location of IceCube at the South Pole. Its spectral and spatial PDF is determined by the atmospheric neutrinos \cite{2011PhRvD..83a2001A}. Assuming that the source signal provides a minor contribution to the overall count statistics, we calculate $B$ directly from the data, by computing the distribution of detected events in declination and energy. 

The source PDF depends on the assumed shape of the source spectrum. Similarly to previous IceCube source searches \cite{icecube_10yr_paper,2021arXiv210109836I}, we consider the powerlaw spectral models.  The spatial model is taken to be Gaussian with the width determined by the angular reconstruction error of events, $\sigma_i$. The IceCube data release \cite{2021arXiv210109836I} adopts a hypothesis that the point-spread function is circularly symmetric, which is perhaps the case if $\sigma$ is small (degree-scale) so that the instrument response does not vary on the scale of the angular uncertainty. Some of the lower quality events can have $\sigma$ in the range up to ten degrees, the scale on which the detector response exhibits sizeable variations. We exclude such events from our analysis by imposing a quality cut $\sigma<1^\circ$. We have verified that such a cut does not degrade the outcomes of the analysis. 

\begin{figure}
    \includegraphics[width=\linewidth]{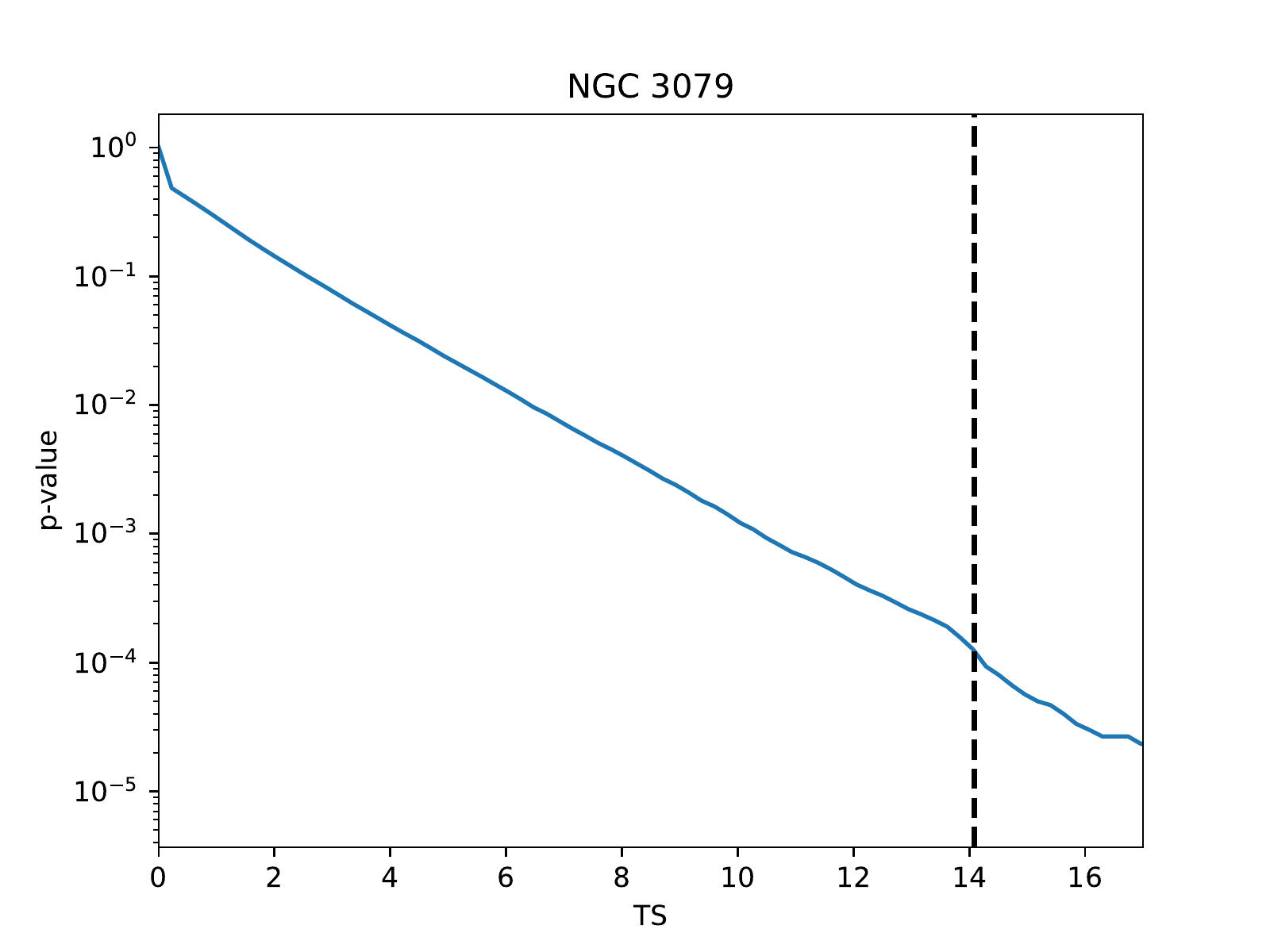}
    \includegraphics[width=\linewidth]{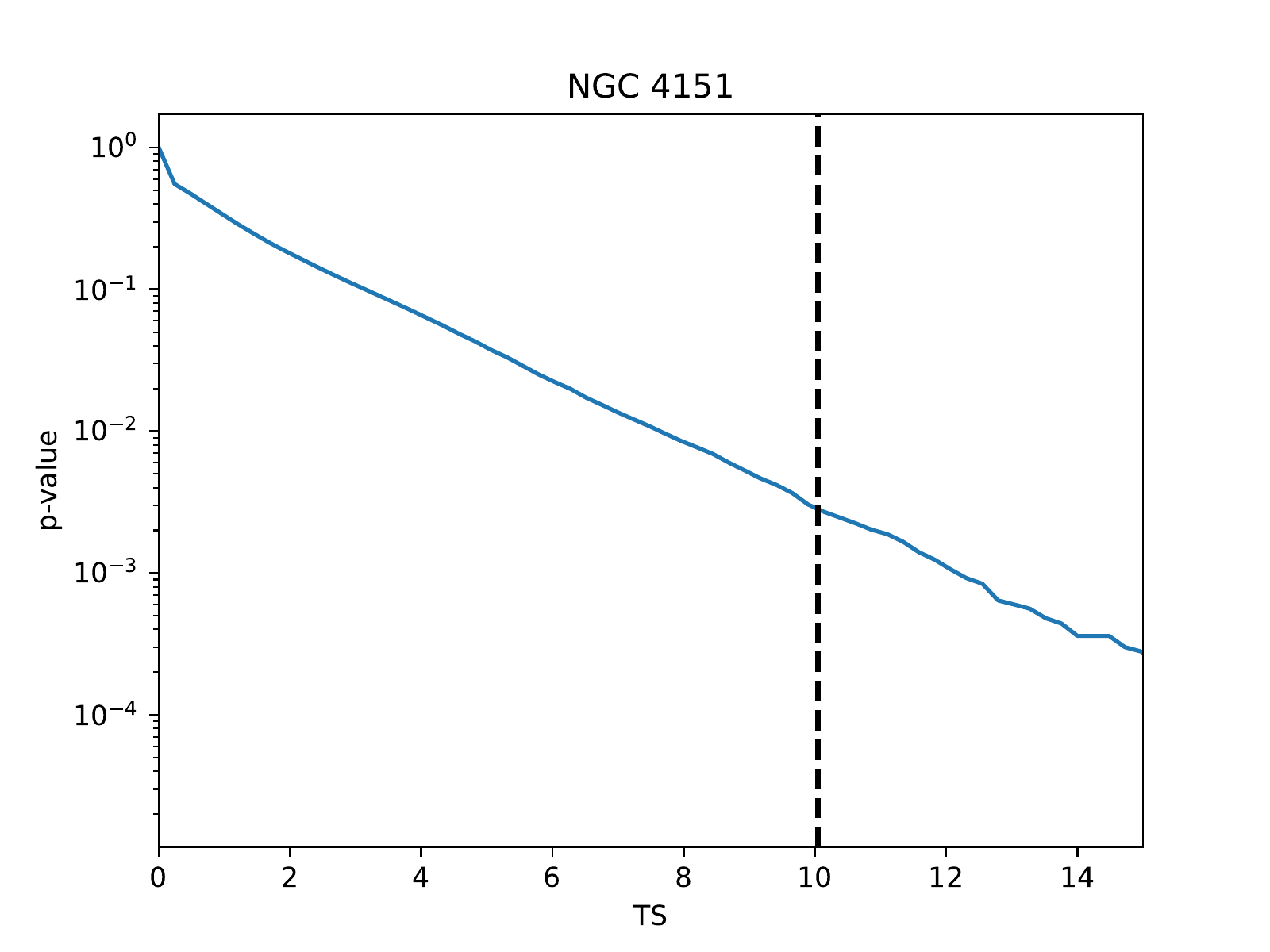}
    \caption{Distribution of TS values found in background fluctuations. Vertical lines show the TS values found for NGC 4151 and NGC 3079.  \textit{Upper panel:} on the position of NGC3079, \textit{lower panel:} on the position of NGC4151.}
    \label{fig:TS_distribution}
\end{figure}

To test the presence of the signal, we calculate the Test Statistic
\begin{equation}
TS(N_s)=2(\log L(N_s)-\log L(0))
\end{equation}
that compares the likelihood of the presence of non-zero signal with any number of counts $N_s$ to the likelihood of the null hypothesis of zero signal counts. To estimate the detection significance of the source we perform Monte-Carlo simulations scrambling the RA of the IceCube events and searching for excesses of TS values allowing for adjustment of the spectral index. Fig. \ref{fig:TS_distribution} shows the result of such simulations. 

To verify our analysis methods, we first reproduce the evidence for the neutrino signal from  NGC 1068. The source is detected with the TS value $TS=22.0$ in the seven-year exposure of the 86 string detector. The data can be fit with the powerlaw spectral model 
\begin{equation}
\frac{dF_{\nu_\mu}}{dE_\nu}=F_{\nu_\mu,0}\left(\frac{E_\nu}{1\mbox{ TeV}}\right)^{-\Gamma_\nu}
\end{equation}
with flux normalization {$F_{\nu_\mu,0}=2.2+3.6-1.1 \times10^{-11}\,\mathrm{TeV^{-1}cm^{-2}s^{-1}}$ and slope $\Gamma_\nu=3.3+0.8-0.3$. 
The corresponding $0.3-100$~TeV neutrino flux is $F_{\nu_\mu,100}=1.6+1.2-0.6\times10^{-10}\,\mathrm{TeV\,cm^{-2}s^{-1}}$.}

\begin{figure}
    \includegraphics[width=\linewidth]{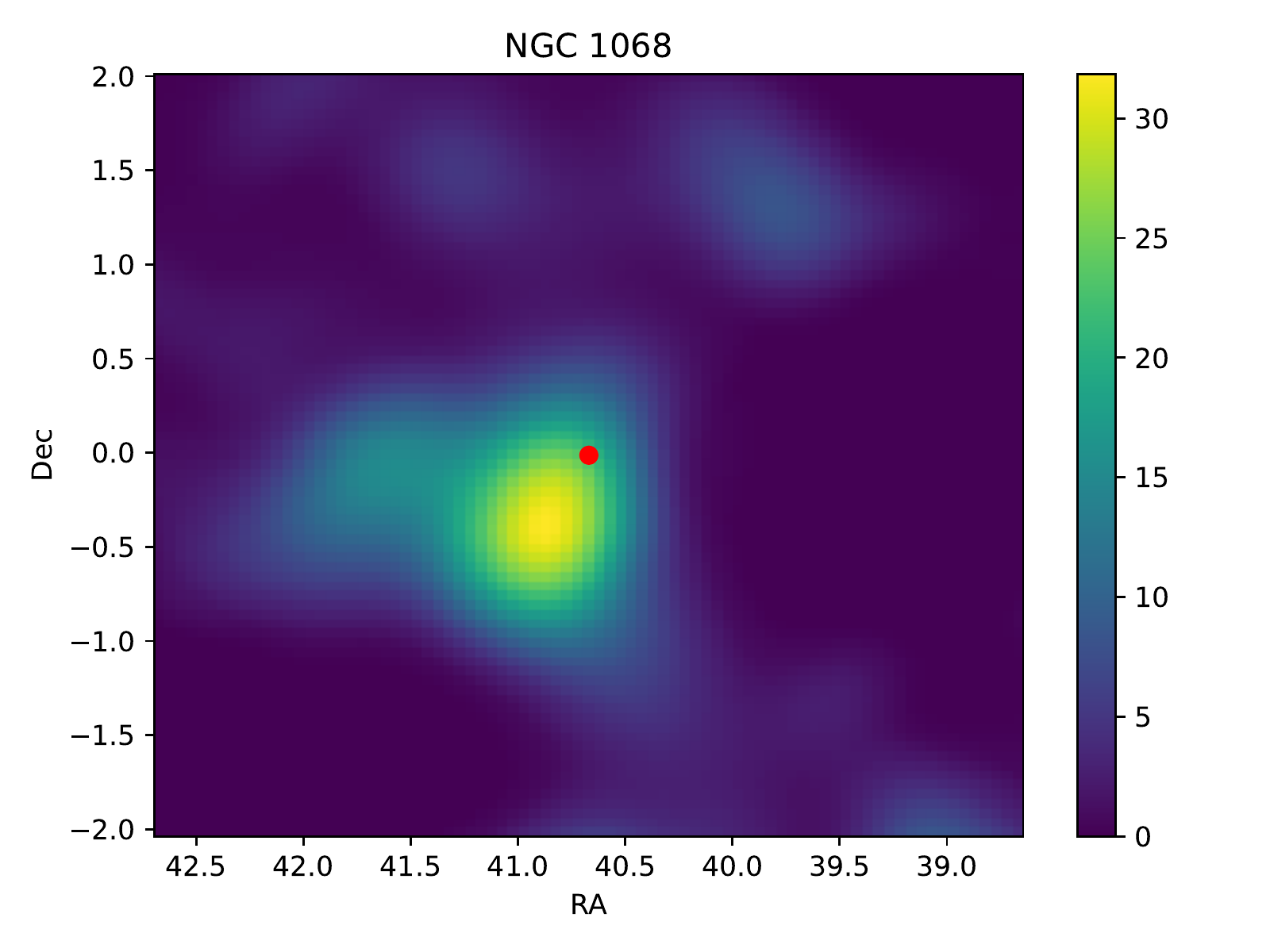}
    \caption{Map of TS values around the positions of NGC 1068. Red dot marks the catalog source position. }
    \label{fig:TS_map_1068}
\end{figure}

Fig. \ref{fig:TS_map_1068} shows the map of TS values. Similar to the analysis of Ref. \cite{2022Sci...378..538I}, we find that  the highest TS position is displaced from the source to the previously reported best-fit position of the IceCube source $RA=40.9^\circ,\  Dec=-0.3^\circ$. {This map, as well as the maps shown in Fig. 2 of the main text, was produced fixing the spectral index $\Gamma$ to its best-fit value.}

\begin{figure}
    \includegraphics[width=\linewidth]{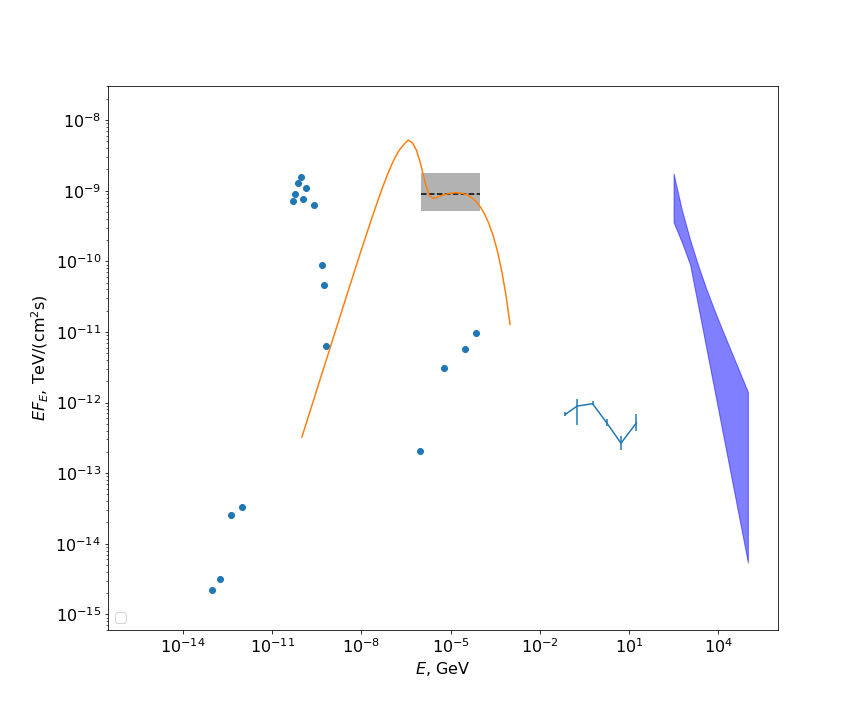}
     \caption{Multi-messenger spectral energy distribution of NGC 1068. BLue data points  are from the central parsec region, from Ref. \cite{2010MNRAS.402..724P}.  Grey horizontal dashed line and gray shading show the estimated flux that would be measured from an un-obscured hard X-ray source\cite{2016MNRAS.456L..94M}. Grey curve shows a spectral model of comptonised emission from an accretion disk with maximal temperature $T_d=100$~eV \cite{1997ASPC..121..101M}. 
     Blue butterfly shows the neutrino spectrum measurement. 
     }
    \label{fig:sed_1068}
\end{figure}

Fig. \ref{fig:sed_1068} shows comparison of the neutrino flux with the electromagnetic flux of NGC 1068. Horizontal dashed line and gray band shows the level of hard X-ray flux of the source in the absence of attenuation by the photoelectric effect and Compton scattering, based on Ref.  \cite{2016MNRAS.456L..94M}. As it is discussed in the main text, softness of the neutrino spectrum suggests that most of the neutrino source power is emitted at the lowest energies accessible to IceCube.  Orange model curve shows a Comptonised accretion disk spectral model of electromagnetic emission from the Seyfert nucleus, from Ref. \cite{1997ASPC..121..101M}. In the absence of obscuration, the overall luminosity of the source is dominated by the luminosity of the accretion disk (with assumed maximum temperature $T_d=100$~eV). One can see that the neutrino source power may be comparable to the overall power of the emission from the accretion, if the neutrino spectrum extends as a soft powerlaw even below the energy threshold of IceCube.

\begin{figure*}
    \includegraphics[width=0.49\linewidth]{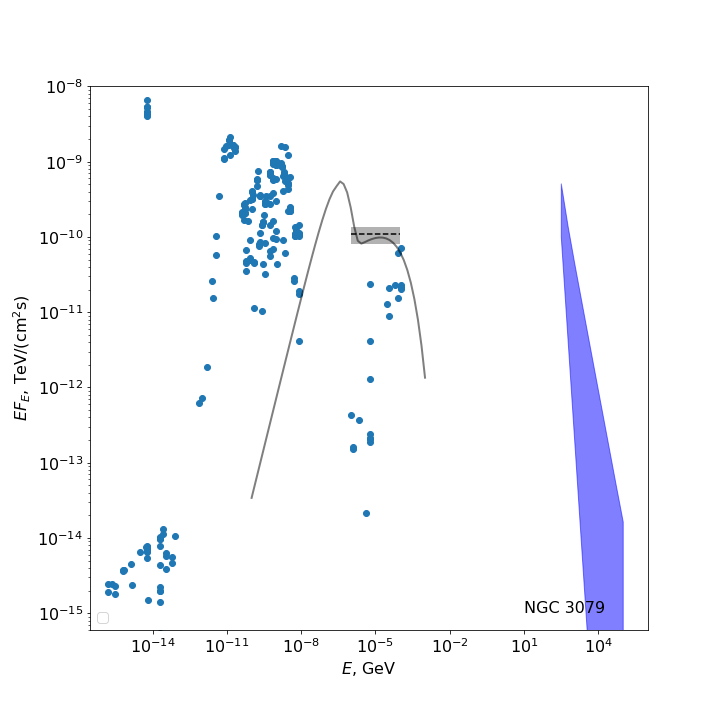}
    \includegraphics[width=0.49\linewidth]{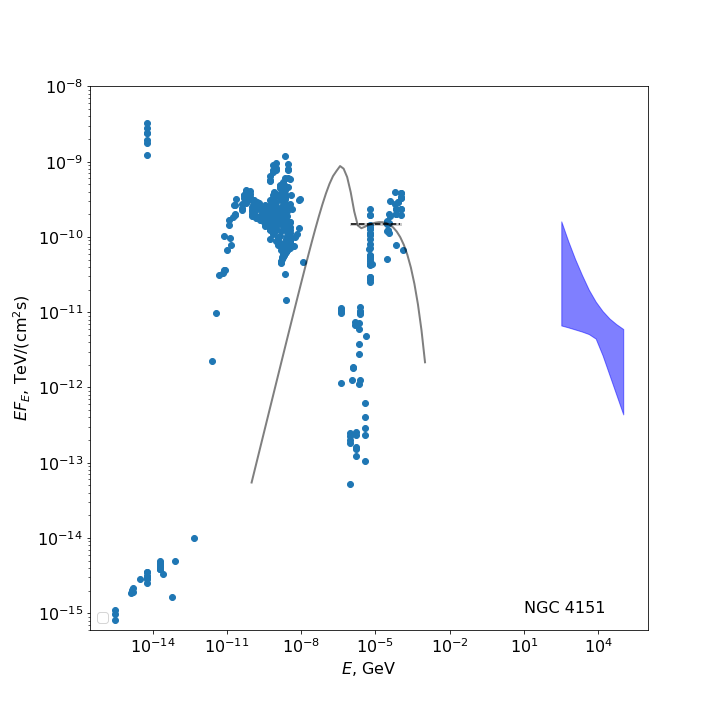}
     \caption{Multi-messenger spectral energy distributions of NGC 3079 and NGC 4151.  Blue data points are NED photometry from {\tt https://ned.ipac.caltech.edu}.  Other notations are the same as in Fig. \ref{fig:sed_1068}. 
     }
    \label{fig:sed_3079}
\end{figure*}

Fig. \ref{fig:sed_3079} shows the spectral energy distributons of the two additional sources, NGC 3079 and NGC 4151. The neutrino spectra of the two sources are also soft, with the low energy emission at the IceCube energy threshold dominating the source power output. Similar to NGC 1068, the neutrino emission power may be comparable to the overall power of the hot corona and of the accretion flow, if the soft spectrum extends even below the energies accessible to IceCube.

\bibliography{references}